\documentclass[12pt]{iopart}
\usepackage{iopams,epsf,psfig}

\newcommand{\half}{\mbox{$\textstyle\frac{1}{2}$}}

\begin{document}
\title[The $\mathcal{C}$ Operator in $\mathcal{PT}$-Symmetric Theories]{The
$\mathcal{C}$ Operator in $\mathcal{PT}$-Symmetric \\  Quantum Theories}

\author[Bender, Brod, Refig, and Reuter]{Carl~M.~Bender\footnote{Permanent
address: Department of Physics, Washington University, St. Louis, MO 63130,
USA.}, Joachim Brod, Andr\'e Refig,\\ and Moritz Reuter}

\address{Blackett Laboratory, Imperial College, London SW7 2BZ, UK}

\begin{abstract}
The Hamiltonian $H$ specifies the energy levels and the time evolution of a
quantum theory. It is an axiom of quantum mechanics that $H$ be Hermitian
because Hermiticity guarantees that the energy spectrum is real and that the
time evolution is unitary (probability preserving). This paper investigates an
alternative way to construct quantum theories in which the conventional
requirement of Hermiticity (combined transpose and complex conjugate) is
replaced by the more physically transparent condition of space-time reflection
($\mathcal{PT}$) symmetry. It is shown that if the $\mathcal{PT}$ symmetry of a
Hamiltonian $H$ is not broken, then the spectrum of $H$ is real. Examples of
$\mathcal{PT}$-symmetric non-Hermitian quantum-mechanical Hamiltonians are
$H=p^2+ix^3$ and $H=p^2-x^4$. The crucial question is whether $\mathcal{P
T}$-symmetric Hamiltonians specify physically acceptable quantum theories in
which the norms of states are positive and the time evolution is unitary. The
answer is that a Hamiltonian that has an unbroken $\mathcal{PT}$ symmetry also
possesses a physical symmetry represented by a linear operator called $\mathcal{
C}$. Using $\mathcal{C}$ it is shown how to construct an inner product whose
associated norm {\it is} positive definite. The result is a new class of fully
consistent complex quantum theories. Observables are defined, probabilities are
positive, and the dynamics is governed by unitary time evolution. After a review
of $\mathcal{P T}$-symmetric quantum mechanics, new results are presented here
in which the $\mathcal{C}$ operator is calculated perturbatively in quantum
mechanical theories having several degrees of freedom.
\end{abstract}

\submitto{\JPA}

\section{Introduction}
\label{s1}
In this paper we present a brief review of some recent work on an alternative
way to formulate of quantum mechanical models and present new results concerning
the perturbative calculation of what has become known in this theory as the 
$\mathcal{C}$ operator.

In the conventional formulation of quantum mechanics the Hamiltonian $H$, which
incorporates the symmetries and specifies the dynamics of a quantum theory, must
be Hermitian: $H=H^\dagger$. The usual meaning of the symbol $\dagger$, which
indicates Dirac Hermitian conjugation, is combined transpose and complex
conjugation. It is commonly thought that a Hamiltonian must be Hermitian in
order to ensure that the energy spectrum (the eigenvalues of $H$) is real and
that the time evolution of the theory is unitary (probability is conserved in
time). Although $H=H^\dagger$ is sufficient to guarantee these properties, it is
not necessary. Indeed, we believe that this condition of Hermiticity is a
mathematical requirement whose physical basis is somewhat obscure. Recently, a
more physical alternative axiom called space-time reflection symmetry ($\mathcal
{PT}$ symmetry), $H=H^\mathcal{PT}$, has been investigated. This symmetry allows
for the possibility of complex non-Hermitian Hamiltonians but still leads to a
consistent theory of quantum mechanics.

Because $\mathcal{PT}$ symmetry is an alternative condition to conventional
Hermiticity, it is now possible to construct infinitely many new Hamiltonians
that would have been rejected in the past because they are not Hermitian in the
usual sense. One example of such a Hamiltonian is $H=\half p^2+\half\mu^2x^2+i
\epsilon x^3$, which is the quantum mechanical analog of the quantum field
theoretic Hamiltonian $H=\int d{\bf x}\left[\half\pi^2+\half(\nabla\varphi)^2+
\half\mu^2\varphi^2+i\epsilon\varphi^3\right]$. Another example of a $\mathcal{P
T}$-symmetric non-Hermitian Hamiltonian is $H=\half p^2+\half\mu^2 x^2-\epsilon
x^4$, which is the $\mathcal{PT}$-symmetric analog of the quantum field
theoretic Hamiltonian $H=\int d{\bf x}\left[\half\pi^2+\half(\nabla\varphi)^2+
\half\mu^2\varphi^2-\epsilon\varphi^4\right]$. This latter Hamiltonian could be
an interesting candidate for describing the Higgs sector of the standard model.
It should be emphasized that we do not regard the condition of conventional
Hermiticity as wrong. Rather, we view the condition of $\mathcal{PT}$ symmetry
as offering the possibility of studying new kinds of quantum theories that have
heretofore never been studied because they have been thought to be physically
unacceptable.

Let us review the properties of the space reflection (parity) operator $\mathcal
{P}$ and the time-reflection operator $\mathcal{T}$: $\mathcal{P}$ is a {\it
linear} operator with the property that $\mathcal{P}^2=1$ and has the effect $p
\to-p$ and $x\to-x$; $\mathcal{T}$ is an antilinear operator with the property
that $\mathcal{T}^2=1$ and has the effect $p\to-p$, $x\to x$, and $i\to-i$. The
operator $\mathcal{T}$ is called {\it antilinear} because it changes the sign of
$i$. We know that it reverses the sign of $i$ because, like $\mathcal{P}$, this
operator preserves the fundamental commutation relation of quantum mechanics,
$[x,p]=i$, known as the Heisenberg algebra.

It is easy to construct Hamiltonians of the form $H=p^2+V(x)$ that are not
Hermitian but do possess $\mathcal{PT}$ symmetry. The trick is to take the
potential to be a function of $ix$: $V=V(ix)$. We also impose a general
condition that has not been widely emphasized in the literature; namely, we
require that $H$ be symmetric: $H=H^{\rm T}$, where ${\rm T}$ represents the
transpose. The reason for this symmetry condition will become clear later on.
For example, consider the one-parameter family of symmetric Hamiltonians
\begin{eqnarray}
H=p^2+x^2(ix)^\epsilon,
\label{e1}
\end{eqnarray}
where $\epsilon$ is real. While $H$ in (\ref{e1}) is not symmetric under
$\mathcal{P}$ or $\mathcal{T}$ separately, it is invariant under their combined
operation. Such Hamiltonians are said to possess space-time reflection symmetry.
Other examples of complex Hamiltonians having $\mathcal{PT}$ symmetry are
$H=p^2+x^4(ix)^\epsilon$, $H=p^2+x^6(ix)^\epsilon$, and so on \cite{rr1}. Note
that these classes of Hamiltonians are all {\it different}. For example, the
Hamiltonian obtained by continuing $H$ in (\ref{e1}) along the path $\epsilon:\,
0\to8$ has a different spectrum from the Hamiltonian that is obtained by
continuing $H=p^2+x^6(ix)^\epsilon$ along the path $\epsilon:\,0\to4$. This is
because the boundary conditions on the eigenfunctions are different.

The class of $\mathcal{PT}$-symmetric Hamiltonians is larger than and includes
real symmetric Hermitians because any real symmetric Hamiltonian is
$\mathcal{PT}$-symmetric. For example, consider the real symmetric Hamiltonian
$H=p^2+x^2+2x$. This Hamiltonian is time-reversal symmetric, but according to
the usual definition of space reflection for which $x\to-x$, this Hamiltonian
does not appear to have $\mathcal{PT}$ symmetry. However, the parity operator is
defined only up to unitary equivalence, and if we express the Hamiltonian in the
form $H=p^2+(x+1)^2-1$, then it is evident that $H$ is $\mathcal{PT}$ symmetric
provided that the parity operator performs a space reflection about the point
$x=-1$ rather than $x=0$. See Ref. \cite{rr2} for the general construction of
the relevant parity operator.

\begin{figure}[t]
\vspace{3.85in}
\includegraphics{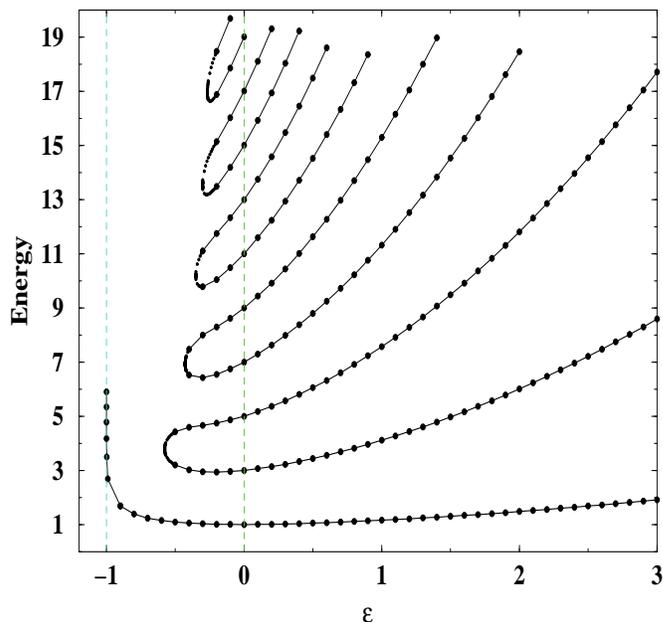}
\vspace{-17mm}
\caption{Energy levels of the Hamiltonian $H=p^2+x^2(ix)^\epsilon$ as a function
of the parameter $\epsilon$. There are three regions: When $\epsilon\geq0$, the
spectrum is real and positive and the energy levels rise with increasing
$\epsilon$. The lower bound of this region, $\epsilon=0$, corresponds to the
harmonic oscillator, whose energy levels are $E_n=2n+1$. When $-1<\epsilon<0$,
there are a finite number of real positive eigenvalues and an infinite number of
complex conjugate pairs of eigenvalues. As $\epsilon$ decreases from $0$ to
$-1$, the number of real eigenvalues decreases; when $\epsilon\leq-0.57793$, the
only real eigenvalue is the ground-state energy. As $\epsilon$ approaches
$-1^+$, the ground-state energy becomes infinite. When $\epsilon\leq-1$ there
are no real eigenvalues.}
\label{f1}
\end{figure}

With properly defined boundary conditions the spectrum of the Hamiltonian $H$ in
(\ref{e1}) is {\em real and positive} when $\epsilon\geq0$ \cite{rr3} and the
spectrum is partly real and partly complex when $\epsilon<0$. The eigenvalues
have been computed numerically to very high precision, and the real eigenvalues
are plotted as functions of $\epsilon$ in Fig.~\ref{f1}.

The $\mathcal{PT}$ symmetry of a Hamiltonian $H$ is said to be {\it unbroken} if
all of the eigenfunctions of $H$ are simultaneously eigenfunctions of $\mathcal{
PT}$. Note that even if a system is defined by an equation that possesses a
discrete symmetry, the solution to this equation need not exhibit that symmetry.
For example, although the differential equation ${\ddot y}(t)=y(t)$ is symmetric
under time reversal $t\to-t$, the solutions $y(t)=e^t$ and $y(t)=e^{-t}$ do not
exhibit time-reversal symmetry; other solutions, such as $y(t)=\cosh(t)$, are
time-reversal symmetric. The same is true of a system whose Hamiltonian is
$\mathcal{PT}$ symmetric. Even if the Schr\"odinger equation and corresponding
boundary conditions are $\mathcal{PT}$ symmetric, the wave function that solves
the Schr\"odinger equation boundary value problem need not be symmetric under
space-time reflection. When the solution exhibits $\mathcal{PT}$ symmetry, we
say that the $\mathcal{PT}$ symmetry is unbroken. Conversely, if the solution
does not possess $\mathcal{PT}$ symmetry, we say that the $\mathcal{PT}$
symmetry is broken.

It is extremely easy to prove that if the $\mathcal{PT}$ symmetry of a
Hamiltonian $H$ is unbroken, then the spectrum of $H$ is real: Assume that (i)
$H$ possesses $\mathcal{PT}$ symmetry (that is, that $H$ commutes with the
$\mathcal{PT}$ operator), and (ii) if $\phi$ is an eigenstate of $H$ with
eigenvalue $E$, then it is simultaneously an eigenstate of $\mathcal{PT}$ with
eigenvalue $\lambda$ (it is this second assumption of {\it unbroken} $\mathcal{P
T}$ symmetry that is crucial):
\begin{eqnarray}
H\phi=E\phi\quad{\rm and}\quad\mathcal{PT}\phi=\lambda\phi.
\label{e2}
\end{eqnarray}
We first show that the eigenvalue $\lambda$ is a pure phase. We multiply
$\mathcal{PT}\phi=\lambda\phi$ on the left by $\mathcal{PT}$ and use the fact
that $\mathcal{P}$ and $\mathcal{T}$ commute and that $\mathcal{P}^2=\mathcal{
T}^2=1$ to conclude that $\phi=\lambda^*\lambda\phi$ and thus $\lambda=e^{i
\alpha}$ for some real $\alpha$. Next, we introduce a convention used throughout
this paper. Without loss of generality we replace the eigenstate $\phi$ by $e^{
-i\alpha/2}\phi$ so that its eigenvalue under the operator $\mathcal{PT}$ is
unity: $\mathcal{PT}\phi=\phi$. Next, we multiply the eigenvalue equation $H
\phi=E\phi$ on the left by $\mathcal{PT}$ and use $[\mathcal{PT},H]=0$ to obtain
$E\phi=E^*\phi$. Hence, $E=E^*$ and the eigenvalue $E$ is real.

The crucial step in the argument above is the assumption that $\phi$ is
simultaneously an eigenstate of $H$ and $\mathcal{PT}$. In quantum mechanics if
a linear operator $X$ commutes with the Hamiltonian $H$, then the eigenstates of
$H$ are also eigenstates of $X$. However, we emphasize that the operator
$\mathcal{PT}$ is not linear (it is antilinear) and thus we must make the extra
assumption that the $\mathcal{PT}$ symmetry of $H$ is unbroken; that is, that
$\phi$ is simultaneously an eigenstate of $H$ and $\mathcal{PT}$. This extra
assumption is nontrivial because it is difficult to determine {\it a priori}
whether the $\mathcal{PT}$ symmetry of a particular Hamiltonian $H$ is broken or
unbroken. For the Hamiltonian $H$ in (\ref{e1}) the $\mathcal{PT}$ symmetry is
unbroken when $\epsilon\geq0$ and it is broken when $\epsilon<0$. The
conventional Hermitian Hamiltonian for the quantum mechanical harmonic
oscillator lies at the boundary of the unbroken and the broken regimes.
Recently, Dorey {\em et al.} proved rigorously that the spectrum of $H$ in
(\ref{e1}) is real and positive \cite{rr4} in the region $\epsilon\geq0$. Many
other $\mathcal{PT}$-symmetric Hamiltonians for which space-time reflection
symmetry is not broken have been investigated, and the spectra of these
Hamiltonians have also been shown to be real and positive \cite{rr5}.

It is important to know whether a given non-Hermitian $\mathcal{PT}$-symmetric
Hamiltonian has a positive real spectrum, but the most urgent question is
whether such a Hamiltonian defines a physical theory of quantum mechanics. By a
{\it physical theory} we mean that there is a Hilbert space of state vectors and
that this Hilbert space has an inner product with a positive norm. In quantum
mechanics we interpret the norm of a state as a probability and this probability
must be positive. Furthermore, we must show that the time evolution of the
theory is unitary. This means that as a state vector evolves in time the
probability does not leak away. With these considerations in mind one would
wonder whether a Hamiltonian such as $H$ in (\ref{e1}) gives a consistent
quantum theory. Indeed, early investigations of this Hamiltonian have shown that
while the spectrum is entirely real and positive when $\epsilon\geq0$, one
inevitably encountered the severe problem of a Hilbert space endowed with an
indefinite metric \cite{rr6}.

However, there is a new symmetry that all $\mathcal{PT}$-symmetric Hamiltonians
having an unbroken $\mathcal{PT}$-symmetry possess \cite{rr7}. We denote the
operator representing this symmetry by $\mathcal{C}$ because the properties of
this operator resemble those of the charge conjugation operator in particle
physics. This allows us to introduce an inner product structure associated with
$\mathcal{CPT}$ conjugation for which the norms of quantum states are positive
definite. Because of this we can say that $\mathcal{PT}$ symmetry is an
alternative to conventional Hermiticity; it introduces the new concept of a {\sl
dynamically determined} inner product (one that is defined by the Hamiltonian
itself). Consequently, we can extend the Hamiltonian and its eigenstates into
the complex domain so that the associated eigenvalues are real and the
underlying dynamics is unitary. This shows that $\mathcal{PT}$-symmetric
Hamiltonians are Hermitian in an extended (non-Dirac) sense.

This paper is organized as follows. In Sec.~\ref{s2} we give a general
discussion of the $\mathcal{C}$ operator and in Sec.~\ref{s3} we present a
simple $2\times2$ matrix example of this operator. In Sec.~\ref{s4} we show
how to calculate $\mathcal{C}$ using perturbation theory for the cubic
Hamiltonian $H=\half p^2+\half\mu^2x^2+i\epsilon x^3$. In Secs.~\ref{s5} and
\ref{s6} we calculate $\mathcal{C}$ for quantum mechanical Hamiltonians having
two and three degrees of freedom. This is the principal new result in this
paper. In Secs.~\ref{s7} and \ref{s8} we consider possible physical applications
and draw some conclusions.

\section{Construction of the $\mathcal{C}$ Operator}
\label{s2}
We begin by summarizing the mathematical properties of the solution to the
Sturm-Liouville differential equation eigenvalue problem
\begin{equation}
-\phi_n''(x)+x^2(ix)^\epsilon\phi_n(x)=E_n\phi_n(x)
\label{e3}
\end{equation}
associated with the Hamiltonian $H$ in (\ref{e1}). This differential equation
must be imposed on an infinite contour in the complex-$x$ plane. For large $|x|$
this contour lies in wedges placed symmetrically with respect to the
imaginary-$x$ axis \cite{rr3}. The boundary conditions on the eigenfunctions are
that $\phi(x)\to0$ exponentially rapidly as $|x|\to\infty$ along the contour.
For $0\leq\epsilon<2$, the contour may lie on the real axis.

When $\epsilon\geq0$, the Hamiltonian has an unbroken $\mathcal{PT}$ symmetry.
Thus, we know that the eigenfunctions $\phi_n(x)$ are simultaneously eigenstates
of the $\mathcal{PT}$ operator:
\begin{equation}
\mathcal{PT}\phi_n(x)=\lambda_n\phi_n(x).
\label{e4}
\end{equation}
As we argued above, $\lambda_n$ is a pure phase and, without loss of generality,
for each $n$ this phase can be absorbed into $\phi_n(x)$ by a multiplicative
rescaling so that the new eigenvalue under $\mathcal{PT}$ is unity:
\begin{equation}
\mathcal{PT}\phi_n(x)=\phi_n^*(-x)=\phi_n(x).
\label{e5}
\end{equation}

It is not known rigorously yet, but there is strong evidence that when properly
normalized the eigenfunctions $\phi_n(x)$ are complete. The coordinate-space
statement of completeness reads
\begin{equation}
\sum_n(-1)^n\phi_n(x)\phi_n(y)=\delta(x-y)\qquad(x,y~{\rm real}).
\label{e6}
\end{equation}
This nontrivial result has been verified numerically to extremely high accuracy 
(twenty decimal places) \cite{rr8,rr9}. There is a factor of $(-1)^n$ in the
sum. This unusual factor does not appear in conventional quantum mechanics. The
presence of this factor is explained in the following discussion of
orthonormality [see (\ref{e8})] in which we encounter the problem associated
with non-Hermitian $\mathcal{PT}$-symmetric Hamiltonians.

There seems to be a natural choice for the inner product of two functions $f(x)$
and $g(x)$:
\begin{equation}
(f,g)\equiv\int dx\,[\mathcal{PT}f(x)]g(x),
\label{e7}
\end{equation}
where $\mathcal{PT}f(x)=[f(-x)]^*$ and the integration path is the appropriate
contour in the complex-$x$ plane. [We will see that (\ref{e7}) is {\it not} the
correct choice for an inner product because it gives an indefinite metric. The
correct inner product will be defined shortly, but studying this inner product
is useful because it reveals the underlying mathematical structure of the
theory.] The apparent advantage of this inner product is that the associated
norm $(f,f)$ is independent of the overall phase of $f(x)$ and is conserved in
time because $H$ commutes with $\mathcal{PT}$ and the time-evolution operator is
$e^{-iHt}$. Phase independence is desired because in quantum mechanics the
objective is to construct a space of rays to represent quantum mechanical
states. With respect to this inner product, the eigenfunctions $\phi_m(x)$ and
$\phi_n(x)$ of $H$ in (\ref{e1}) are orthogonal for $n\neq m$ because $H$ is
symmetric. However, when we set $m=n$ we see by direct numerical calculation
that the norm is evidently {\it not positive}:
\begin{equation}
(\phi_m,\phi_n)=(-1)^n\delta_{mn}.
\label{e8}
\end{equation}
This result is apparently true for all values of $\epsilon$ in (\ref{e3}), and
it has been verified numerically to extremely high precision. Because the norms
of the eigenfunctions alternate in sign, the metric associated with the
$\mathcal{PT}$ inner product $(\cdot,\cdot)$ is indefinite. This sign
alternation appears to be a {\it generic} feature of this $\mathcal{PT}$ inner
product. [Extensive numerical calculations verify that the formula in (\ref{e8})
holds for all $\epsilon\geq0$.] We emphasize that while the sign of the norm of
$\phi_n$ is hard to verify analytically, the orthogonality of $\phi_m$ and
$\phi_n$ is a trivial consequence of the symmetry of $H$. It is necessary to
assume that $H$ be symmetric in order to have this orthogonality.

In spite of the nonpositivity of the inner product, it is instructive to proceed
with the usual analysis that one would perform for any Sturm-Liouville problem
of the form $H\phi_n=E_n\phi_n$. First, we use the inner product formula
(\ref{e8}) to verify that (\ref{e6}) is the representation of the unity
operator. That is, we verify that $\int dy\,\delta(x-y)\delta(y-z)=\delta(x-z)$.

Second, we show how to reconstruct the parity operator $\mathcal{P}$ in terms of
the eigenstates. In coordinate space the parity operator is given by $\mathcal{P
}(x,y)=\delta(x+y)$, so from (\ref{e6}) we get
\begin{equation}
\mathcal{P}(x,y)=\sum_n(-1)^n\phi_n(x)\phi_n(-y).
\label{e9}
\end{equation}
By virtue of (\ref{e8}) the square of the parity operator is unity: $\mathcal{P}
^2=1$.

Third, we reconstruct $H$ in coordinate space:
\begin{equation}
H(x,y)=\sum_n(-1)^nE_n\phi_n(x)\phi_n(y).
\label{e10}
\end{equation}
Using (\ref{e6}) -- (\ref{e8}) we can see that $H$ satisfies $H\phi_n(x)=E_n
\phi_n(x)$.

Fourth, we construct the coordinate-space Green's function $G(x,y)$:
\begin{equation}
G(x,y)=\sum_n(-1)^n\frac{1}{E_n}\phi_n(x)\phi_n(y).
\label{e11}
\end{equation}
The Green's function is the functional inverse of $H$; that is, $G$ satisfies
\begin{equation}
\int dy\,H(x,y)G(y,z)=\left[-\frac{d^2}{dx^2}+x^2(ix)^\epsilon\right]G(x,z)=
\delta(x-z).
\label{e12}
\end{equation}
The time-independent Schr\"odinger equation (\ref{e3}) cannot be solved
analytically; it can only be solved numerically or perturbatively. However, the
differential equation for $G(x,z)$ in (\ref{e12}) {\it can} be solved exactly
and in closed form because it is a Bessel equation \cite{rr9}. The technique is
to consider the case $0<\epsilon<2$ so that we may treat $x$ as real and then to
decompose the $x$ axis into two regions, $x>z$ and $x<z$. We solve the
differential equation in each region in terms of Bessel functions and patch the
solutions together at $x=z$. Then, using this coordinate-space representation of
the Green's function, we construct an exact closed-form expression for the {\it
spectral zeta function} (sum of the inverses of the energy eigenvalues). To do
so we set $z=x$ in $G(x,z)$ and use (\ref{e8}) to integrate over $x$. For all
$\epsilon>0$ we obtain \cite{rr9}
\begin{equation}
\sum_n {1\over E_n}=\left[1+{\cos\left({3\epsilon\pi\over2\epsilon+8}\right)\sin
\left({\pi\over4+\epsilon}\right)\over\cos\left({\epsilon\pi\over4+2\epsilon}
\right)\sin\left({3\pi\over4+\epsilon}\right)}\right]{\Gamma\left({1\over4+
\epsilon}\right)\Gamma\left({2\over4+\epsilon}\right)\Gamma\left({\epsilon\over
4+\epsilon}\right)\over(4+\epsilon)^{4+2\epsilon\over4+\epsilon}\Gamma\left({1+
\epsilon\over4+\epsilon}\right)\Gamma\left({2+\epsilon\over4+\epsilon}\right)}.
\label{e13}
\end{equation}
This result has been verified to extremely high numerical accuracy \cite{rr9}.

All of these general Sturm-Liouville constructions are completely standard. But
now we must address the crucial question of whether a $\mathcal{PT}$-symmetric
Hamiltonian defines a physically viable quantum mechanics or whether it merely
provides an amusing Sturm-Liouville eigenvalue problem. The apparent difficulty
with formulating a quantum theory is that the vector space of quantum states is
spanned by energy eigenstates, of which half have $\mathcal{PT}$ norm $+1$ and
half have $\mathcal{PT}$ norm $-1$. Because the norm of the states carries a
probabilistic interpretation in standard quantum theory, the existence of an
indefinite metric in (\ref{e8}) seems to be a serious obstacle. The situation
here in which half of the energy eigenstates have positive norm and half have
negative norm is analogous to the problem that Dirac encountered in formulating
the spinor wave equation in relativistic quantum theory \cite{rr10}. Following
Dirac's approach, we attack the problem of an indefinite norm by finding a
physical interpretation for the negative norm states. We claim that in {\it any}
theory having an unbroken $\mathcal{PT}$ symmetry there exists a symmetry of the
Hamiltonian connected with the fact that there are equal numbers of
positive-norm and negative-norm states. To describe this symmetry we construct a
linear operator denoted by $\mathcal{C}$ and represented in position space as a
sum over the energy eigenstates of the Hamiltonian \cite{rr7}:
\begin{equation}
\mathcal{C}(x,y)=\sum_n\phi_n(x)\phi_n(y).
\label{e14}
\end{equation}

The properties of this new operator $\mathcal{C}$ closely resemble those of the
charge conjugation operator in quantum field theory. For example, we can use
(\ref{e6}) -- (\ref{e8}) to verify that the square of $\mathcal{C}$ is unity
($\mathcal{C}^2=1$):
\begin{equation}
\int dy\,\mathcal{C}(x,y)\mathcal{C}(y,z)=\delta(x-z).
\label{e15}
\end{equation}
Thus, the eigenvalues of $\mathcal{C}$ are $\pm1$. Also, $\mathcal{C}$ commutes
with the Hamiltonian $H$. Therefore, since $\mathcal{C}$ is linear, the
eigenstates of $H$ have definite values of $\mathcal{C}$. Specifically, if the
energy eigenstates satisfy (\ref{e8}), then we have $\mathcal{C}\phi_n=(-1)^n
\phi_n$ because
\begin{equation}
\mathcal{C}\phi_n(x)=\int dy\,\mathcal{C}(x,y)\phi_n(y)=\sum_m\phi_m(x)\int dy\,
\phi_m(y)\phi_n(y).
\label{e16}
\end{equation}
We then use $\int dy\,\phi_m(y)\phi_n(y)=(\phi_m,\phi_n)$ according to our
convention. We conclude that $\mathcal{C}$ is the operator observable that
represents the measurement of the signature of the $\mathcal{PT}$ norm of a
state. Note that since the $\mathcal{C}$ operator measures the $\mathcal{PT}$
norm of a state, we can think of the $\mathcal{PT}$ norm as the $\mathcal{C}$
``charge'' of the state.

The operators $\mathcal{P}$ and $\mathcal{C}$ are distinct square roots of the
unity operator $\delta(x-y)$. That is, $\mathcal{P}^2=\mathcal{C}^2=1$, but
$\mathcal{P}\neq\mathcal{C}$. Indeed, $\mathcal{P}$ is real, while $\mathcal{C}$
is complex. Note that the parity operator in coordinate space is explicitly real
$\mathcal{P}(x,y)=\delta(x+y)$; the operator $\mathcal{C}(x,y)$ is complex
because it is a sum of products of complex functions, as we see in (\ref{e14}).
The complexity of the $\mathcal{C}$ operator can be seen explicitly in
perturbative calculations of $\mathcal{C}(x,y)$ \cite{rr11}. We show how to
perform these perturbative calculations in Secs.~\ref{s4} and \ref{s5}.
Furthermore, these two operators do not commute; in the position representation
\begin{equation}
(\mathcal{CP})(x,y)=\sum_n\phi_n(x)\phi_n(-y)~~{\rm but}~~(\mathcal{PC})(x,y)=
\sum_n\phi_n(-x)\phi_n(y),
\label{e17}
\end{equation}
which shows that $\mathcal{CP}=(\mathcal{PC})^*$. However, $\mathcal{C}$
{\it does} commute with $\mathcal{PT}$.

Finally, having obtained the operator $\mathcal{C}$ we define a new inner
product structure having {\it positive definite} signature by
\begin{equation}
\langle f|g\rangle\equiv\int_{\rm{C}} dx\,[\mathcal{CPT}f(x)]g(x).
\label{e18}
\end{equation}
This inner product is phase independent and conserved in time like the $\mathcal
{PT}$ inner product (\ref{e7}). This is because the time evolution operator,
just as in ordinary quantum mechanics, is $e^{iHt}$. The fact that $H$ commutes
with the $\mathcal{PT}$ and the $\mathcal{CPT}$ operators implies that both
inner products, (\ref{e7}) and (\ref{e18}), remain time independent as the
states evolve in time. However, unlike (\ref{e7}), the inner product (\ref{e18})
is positive definite because $\mathcal{C}$ contributes $-1$ when it acts on
states with negative $\mathcal{PT}$ norm. In terms of the $\mathcal{CPT}$
conjugate, the completeness condition (\ref{e6}) reads
\begin{equation}
\sum_n \phi_n(x)[\mathcal{CPT}\phi_n(y)]=\delta(x-y).
\label{e19}
\end{equation}
Unlike the inner product of conventional quantum mechanics, the $\mathcal{CPT}$
inner product (\ref{e19}) is {\it dynamically determined}; it depends implicitly
on the Hamiltonian.

The operator $\mathcal{C}$ does not exist as a distinct entity in conventional
quantum mechanics. Indeed, if we allow the parameter $\epsilon$ in (\ref{e1}) to
tend to zero, the operator $\mathcal{C}$ in this limit becomes identical with
$\mathcal{P}$. Thus, in this limit the $\mathcal{CPT}$ operator becomes
$\mathcal{T}$, which is just complex conjugation. As a consequence, the inner
product (\ref{e18}) defined with respect to the $\mathcal{CPT}$ conjugation
reduces to the complex conjugate inner product of conventional quantum mechanics
when $\epsilon\to0$. Similarly, in the $\epsilon\to0$ limit, (\ref{e19}) reduces
to the usual statement of completeness $\sum_n\phi_n(x)\phi_n^*(y)=\delta(x-y)$.

The $\mathcal{CPT}$ inner-product (\ref{e18}) is independent of the choice of
integration contour ${\rm C}$ so long as ${\rm C}$ lies inside the asymptotic
wedges associated with the boundary conditions for the Sturm-Liouville problem
(\ref{e2}). Path independence follows from Cauchy's theorem and the analyticity
of the integrand. In ordinary quantum mechanics, where the positive-definite
inner product has the form $\int dx\,f^*(x)g(x)$, the integral must be taken
along the real axis and the path of the integration cannot be deformed into the
complex plane because the integrand is not analytic. [Note that if a function
satisfies a linear ordinary differential equation, then the function is analytic
wherever the coefficient functions of the differential equation are analytic.
The Schr\"odinger equation (\ref{e3}) is linear and its coefficients are
analytic except for a branch cut at the origin; this branch cut can be taken to
run up the imaginary axis. We can choose the integration contour for the inner
product (\ref{e8}) so that it does not cross the positive imaginary axis. Path
independence occurs because the integrand of the inner product (\ref{e8}) is a
product of analytic functions.] The $\mathcal{PT}$ inner product (\ref{e7})
shares with (\ref{e18}) the advantage of analyticity and path independence, but
suffers from nonpositivity. We find it surprising that a positive-definite
metric can be constructed using $\mathcal{CPT}$ conjugation without disturbing
the path independence of the inner-product integral.

Why are $\mathcal{PT}$-symmetric theories unitary? Time evolution is determined
by the operator $e^{-iHt}$ whether the theory is expressed in terms of a
$\mathcal{PT}$-symmetric Hamiltonian or just an ordinary Hermitian Hamiltonian.
To establish the global unitarity of a theory we must show that as a state
vector evolves, its norm does not change in time. If $\psi_0(x)$ is a prescribed
initial wave function belonging to the Hilbert space spanned by the energy
eigenstates, then it evolves into the state $\psi_t(x)$ at time $t$ according to
\begin{equation}
\psi_t(x)=e^{-iHt}\psi_0(x).
\label{e19a}
\end{equation}
With respect to the $\mathcal{CPT}$ inner product defined in (\ref{e18}), the
norm of the vector $\psi_t(x)$ does not change in time, $\langle\psi_t|\psi_t
\rangle=\langle\psi_0|\psi_0\rangle$, because the Hamiltonian $H$ commutes with
the $\mathcal{CPT}$ operator.

Establishing unitarity at a local level is more subtle. Here, we must show that
in coordinate space, there exists a local probability density that satisfies a
continuity equation so that the probability does not leak away. This is a
nontrivial consideration because the probability current flows about in the
complex plane rather than along the real axis as in conventional Hermitian
quantum mechanics. Preliminary numerical studies indeed indicate that the
continuity equation is fulfilled \cite{rr12}.

Just as states in the Schr\"odinger picture evolve in time according to the
usual equation (\ref{e19a}), operators $\mathcal{A}$ also evolve according to
the conventional Heisenberg-picture equation
\begin{equation}
\mathcal{A}(t)=e^{-iHt}\mathcal{A}(0)\,e^{iHt}.
\label{e19b}
\end{equation}
Given this equation, it is clear how to define an {\it observable} in $\mathcal{
PT}$-symmetric quantum mechanics. The crucial property of an observable in any
theory of quantum mechanics is that its expectation value in a state must be
real. This will be true if 
\begin{equation}
\mathcal{A}^{\rm T}=\mathcal{A}^{\mathcal{CPT}}=\mathcal{CPT}\mathcal{A}
\,\,\mathcal{CPT},
\label{e19c}
\end{equation}
where ${\rm T}$ signifies transpose. If this condition is fulfulled by a linear
operator in a theory defined by a $\mathcal{PT}$-symmetric Hamiltonian, we say
that the operator is ``$\mathcal{PCT}$ symmetric'' and is an observable. Note
that this condition is the analog of the usual condition in conventional quantum
mechanics that an observable be Hermitian in the usual Dirac sense $\mathcal{A}^
{\dagger}=\mathcal{A}$:
\begin{equation}
\mathcal{A}^{\rm T}=\mathcal{A}^*.
\label{e19d}
\end{equation}
The condition for an operator to {\it remain} an operator as time evolves is
simply that $H$ be symmetric: $H^{\rm T}=H$. This symmetry condition has been
implicitly assumed in all of the $\mathcal{PT}$-symmetric models discussed in
the literature. Recall that the symmetry of $H$ ensures that eigenstates of $H$
corresponding to different energies will be orthogonal. Note that there are
two time-independent observables in the theory; namely, $H$ and $\mathcal{C}$.

\section{Illustrative Example: A $2\times2$ Matrix Hamiltonian}
\label{s3}

We illustrate the above results concerning $\mathcal{PT}$-symmetric quantum
mechanics using the finite-dimensional $2\times2$ symmetric matrix Hamiltonian
\begin{equation}
H=\left(\begin{array}{cc} re^{i\theta} & s \cr s & re^{-i\theta}
\end{array}\right),
\label{e21}
\end{equation}
where the three parameters $r$, $s$, and $\theta$ are real. This Hamiltonian is
not Hermitian in the usual Dirac sense, but it is $\mathcal{PT}$ symmetric,
where the parity operator is \cite{rr13}
\begin{equation}
\mathcal{P}=\left(\begin{array}{cc} 0 & 1 \cr 1 & 0
\end{array}\right)
\label{e20}
\end{equation}
and $\mathcal{T}$ performs complex conjugation. (Note that $\mathcal{T}$ does
{\it not} perform Hermitian conjugation, or else it would not leave the
commutation relation $[x,p]=i$ invariant.)

There are two parametric regions for this Hamiltonian. When $s^2<r^2\sin^2
\theta$, the energy eigenvalues form a complex conjugate pair. This is the
region of broken $\mathcal{PT}$ symmetry. On the other hand, if $s^2\geq r^2
\sin^2\theta$, then the eigenvalues $\varepsilon_\pm=r\cos\theta\pm\sqrt{s^2-r^2
\sin^2\theta}$ are real. This is the region of unbroken $\mathcal{PT}$ symmetry.
In the unbroken region the simultaneous eigenstates of the operators $H$ and
$\mathcal{PT}$ are
\begin{equation}
|\varepsilon_+\rangle=\frac{1}{\sqrt{2\cos\alpha}}
\left(\begin{array}{c}e^{i\alpha/2}\cr e^{-i\alpha/2}\end{array}\right)
\quad{\rm and}\quad|\varepsilon_-\rangle=\frac{i}{\sqrt{2\cos\alpha}}
\left(\begin{array}{c}e^{-i\alpha/2}\cr-e^{i\alpha/2}\end{array}\right),
\label{e22}
\end{equation}
where we set $\sin\alpha =(r/s)\,\sin\theta$. It is easily verified that
$(\varepsilon_{\pm},\varepsilon_{\pm})=\pm1$ and that $(\varepsilon_{\pm},
\varepsilon_{\mp})=0$, recalling that $(u,v)=(\mathcal{PT}u)\cdot v$. Therefore,
with respect to the $\mathcal{PT}$ inner product, the resulting vector space
spanned by energy eigenstates has a metric of signature $(+,-)$. The condition
$s^2>r^2\sin^2\theta$ ensures that $\mathcal{PT}$ symmetry is not broken. If
this condition is violated, the states (\ref{e22}) are no longer eigenstates of
$\mathcal{PT}$ because $\alpha$ becomes imaginary. (When $\mathcal{PT}$ symmetry
is broken, we find that the $\mathcal{PT}$ norm of the energy eigenstate
vanishes.)

Next, we construct the operator $\mathcal{C}$:
\begin{equation}
\mathcal{C}=\frac{1}{\cos\alpha}
\left(\begin{array}{cc} i\sin\alpha & 1 \cr 1 & -i\sin\alpha \end{array}\right).
\label{e23}
\end{equation}
Note that $\mathcal{C}$ is distinct from $H$ and $\mathcal{P}$ and has the key
property that $\mathcal{C}|\varepsilon_{\pm}\rangle=\pm|\varepsilon_{\pm}
\rangle$. The operator $\mathcal{C}$ commutes with $H$ and satisfies $\mathcal{C
}^2=1$. The eigenvalues of $\mathcal{C}$ are precisely the signs of the
$\mathcal{PT}$ norms of the corresponding eigenstates. Using $\mathcal{C}$ we
construct the new inner product structure $\langle u|v\rangle=(\mathcal{CPT}u)
\cdot v$. This inner product is positive definite because $\langle\varepsilon_{
\pm}|\varepsilon_{\pm}\rangle=1$. Thus, the two-dimensional Hilbert space
spanned by $|\varepsilon_\pm\rangle$, with inner product $\langle\cdot|\cdot
\rangle$, has a Hermitian structure with signature $(+,+)$.

Let us demonstrate explicitly that the $\mathcal{CPT}$ norm of any vector is
positive. For the arbitrary vector $\psi=\left({a\atop b}\right)$, where $a$ and
$b$ are any complex numbers, we see that $\mathcal{T}\psi=\left(a^*\atop b^*
\right)$, that $\mathcal{PT}\psi=\left(b^*\atop a^*\right)$, and that $\mathcal{
CPT}\psi={1\over\cos\alpha}\,\left(a^*+ib^*\sin\alpha\atop b^*-ia^*\sin\alpha
\right)$. Thus, $\langle\psi|\psi\rangle=(\mathcal{CPT}\psi)\cdot\psi={1\over
\cos\alpha}[a^*a+b^*b+i(b^*b-a^*a)\sin\alpha]$. Now let $a=x+iy$ and $b=u+iv$,
where $x$, $y$, $u$, and $v$ are real. Then
\begin{equation}
\langle\psi|\psi\rangle=\left(x^2+v^2+2xv\sin\alpha
+y^2+u^2-2yu\sin\alpha\right)/\cos(\alpha),
\label{e26}
\end{equation}
which is explicitly positive and vanishes only if $x=y=u=v=0$.

Since $\langle u|$ denotes the $\mathcal{CPT}$ conjugate of $|u\rangle$, the
completeness condition reads
\begin{equation}
|\varepsilon_+\rangle\langle\varepsilon_+|+|\varepsilon_-\rangle\langle
\varepsilon_-|=\left(\begin{array}{cc} 1 & 0 \cr 0 &
1\end{array}\right).
\label{e27}
\end{equation}
Furthermore, using the $\mathcal{CPT}$ conjugate $\langle\varepsilon_\pm|$, we 
have $\mathcal{C}$ as $\mathcal{C}=|\varepsilon_+\rangle\langle\varepsilon_+|-|
\varepsilon_-\rangle\langle\varepsilon_-|,$ instead of the representation in
(\ref{e14}), which uses the $\mathcal{PT}$ conjugate.

For the two-state system discussed here, if $\theta\to0$, then the Hamiltonian
(\ref{e21}) becomes Hermitian. However, in this limit $\mathcal{C}$ reduces to
the parity operator $\mathcal{P}$. As a consequence, the requirement of
$\mathcal{CPT}$ invariance reduces to the standard condition of Hermiticity for
a symmetric matrix; namely, that $H=H^*$. This is why the hidden symmetry
$\mathcal{C}$ was not noticed previously. The operator $\mathcal{C}$ emerges
only when we extend a real symmetric Hamiltonian into the complex domain.

\section{Perturbative Calculation of the $\mathcal{C}$ Operator
for an $\boldsymbol{ix^3}$ Theory}
\label{s4}

The $\mathcal{C}$ operator can be calculated in some infinite-dimensional 
quantum mechanical models. For an $x^2+ix^3$ potential $\mathcal{C}$ can be
obtained from the summation in (\ref{e14}) using perturbative methods and for an
$x^2-x^4$ potential $\mathcal{C}$ can be calculated using nonperturbative
methods \cite{rr11}. In this paper we focus on perturbative methods for
calculating $\mathcal{C}$.

Let us consider the $\mathcal{PT}$-symmetric Hamiltonian for a harmonic
oscillator perturbed by an imaginary cubic potential:
\begin{equation}
H=\half p^2+\half x^2+i\epsilon x^3 \quad(\epsilon~{\rm real}).
\label{e27a}
\end{equation}

Following the procedure in Ref.~\cite{rr11}, we note that the energy eigenstates
are solutions of the Schr\"odinger equation
\begin{equation}
H\phi_n(x)=\left(-\half\frac{d^2}{dx^2}+\half x^2+i\epsilon x^3\right)\phi_n(x)
=E_n\phi_n(x).
\label{e27b}
\end{equation}
The eigenstates and the corresponding eigenvalues may be expressed as series in
powers of $\epsilon$ by perturbing around the known energy eigenstates and
eigenvalues of the harmonic oscillator. To second order in perturbation theory
the eigenstates have the form
\begin{equation}
\label{e27c}
\phi_n(x)=\frac{i^n a_n}{\sqrt{2^nn!\sqrt{\pi}}}e^{-\frac{1}{2}x^2}
\Big[H_n(x)+\epsilon P_n(x)+\epsilon^2 Q_n(x)\Big]
\end{equation}
with energies given by
\begin{equation}
E_n=n+\half+\epsilon A_n+\epsilon^2 B_n.
\label{e27d}
\end{equation}
Here, $P_n(x)$ and $Q_n(x)$ are polynomials in $x$ of degree $n+3$ and $n+6$,
respectively; $a_n$ is a normalization constant to be determined. We include a
factor of $ i^n$ because the unperturbed wavefunctions have the form
\begin{equation}
\phi_n^{(0)}(x)=\frac{i^n}{\sqrt{2^nn!\sqrt{\pi}}}e^{-\frac{1}{2}x^2}H_n(x),
\nonumber
\end{equation}
where $H_n(x)$ are Hermite polynomials. This ensures that the unperturbed
wavefunctions are eigenstates of the $\mathcal{PT}$ operator with unit
eigenvalue:
\[ \mathcal{PT}\phi_n^{(0)}(x)=\phi_n^{(0)}(x). \]
The wave functions are $\mathcal{PT}$-normalized according to
\[ \int_{-\infty}^{\infty}dx\left[\phi_n^{(0)}(x)\right]^2=(-1)^n. \]

\subsection{First-Order Calculation of the Energy Eigenstates and Eigenvalues}

Note that to order $\epsilon^0$, (\ref{e27b}) becomes
\begin{equation}
\label{e27e}
\left[-\half\frac{d^2}{dx^2}+x\frac{d}{dx}-n \right]H_n(x)= 0
\end{equation}
and to order $\epsilon^1$, we have
\[\left[-\half\frac{d^2}{dx^2}+\half x^2-n-\half\right]\left(e^{-\frac{1}{2}x^2}
P_n(x)\right)=(A_n-ix^3)e^{-\half x^2}H_n(x),\]
or
\[\left[-\half\frac{d^2}{dx^2}+x\frac{d}{dx}-n\right]P_n(x)=(A_n-ix^3)H_n(x).\]

Since the Hermite polynomials form a complete set, we may rewrite any polynomial
as a linear superposition of Hermite polynomials. Rewriting $P_n(x)$ in this
manner yields
\[\left(-\half\frac{d^2}{dx^2}+x\frac{d}{dx}-n\right)\sum_{k}p_k H_k(x)=(A_n-
i x^3)H_n(x),\]
which with the help of (\ref{e27e}) simplifies to
\begin{equation}
\label{e27f}
\sum_{k}p_k(k-n)H_k(x)=(A_n- i x^3)H_n(x).
\end{equation}
Also, we have
\begin{eqnarray}
x^3H_n(x)&=&\frac{1}{8}H_{n+3}(x)+\frac{3}{4}(n+1)H_{n+1}(x)+\frac{3}{2}n^2
H_{n-1}(x)\nonumber \\
&&\quad +n(n-1)(n-2)H_{n-3}(x).
\label{eq:x3H}
\end{eqnarray}
The coefficient of $H_n(x)$ on the left side of equation (\ref{e27f}) is zero,
and the expression for $x^3H_n(x)$ on the right side does not contain any terms
in $H_n(x)$. Hence, we conclude that $A_n=0$ for all $n$ to first order in
perturbation theory. Thus, the perturbed energy equals the unperturbed energy.

Rewriting equation (\ref{e27f}) as
\begin{eqnarray}
\sum_{k} p_k(k-n)H_k(x) &=& (- i)\bigg[\frac{1}{8}H_{n+3}(x)
+ \frac{3}{4}(n+1)H_{n+1}(x) \nonumber\\
&&+\frac{3}{2}n^2 H_{n-1}(x) +n(n-1)(n-2)H_{n-3}(x)\bigg]
\end{eqnarray}
and comparing coefficients reveals that
\begin{eqnarray}
\label{eq:ppoly}
iP_n(x)&=&\frac{1}{24}H_{n+3}(x)+\frac{3}{4}(n+1)H_{n+1}(x)\nonumber\\
&& -\frac{3}{2}n^2 H_{n-1}(x)-\frac{1}{3}n(n-1)(n-2)H_{n-3}(x).
\end{eqnarray}

\subsection{Second-Order Calculation of the Energy Eigenstates and Eigenvalues}

At order $\epsilon^2$, the eigenproblem becomes
\[ \left(-\half\frac{d^2}{dx^2}+\half x^2-n-\half\right)
\left[e^{-\frac{1}{2}x^2}Q_n(x)\right]
=e^{-\frac{1}{2}x^2}\left[B_nH_n(x)-ix^3P_n(x)\right], \]
or
\[ \left[-\half\frac{d^2}{dx^2}+x\frac{d}{dx}-n\right]Q_n(x)=B_nH_n(x)-ix^3P_n
(x). \]
On posing $Q_n(x)=\sum_k q_kH_k(x)$ this reduces to
\begin{equation}
\label{eq:seriessol1}
\sum_{k} q_k(k-n)H_k(x) = B_n H_n(x)- i x^3 P_n(x).
\end{equation}

Combining equations (\ref{eq:x3H}) and (\ref{eq:ppoly}) we obtain
\begin{eqnarray}
i x^3P_{n}(x)&=& \frac{1}{192}H_{n+6}(x)+\frac{1}{32}(4n+7)H_{n+4}(x)\nonumber\\
&& \hspace{-2cm} +\frac{1}{16}\left(7n^2+33n+27\right)H_{n+2}(x)
+\frac{1}{8}\left(30n^2+30n+11 \right)H_{n}(x) \nonumber \\
&&\hspace{-2cm}
-\frac{1}{4}n(n-1)\left(7n^2-19n+1 \right)H_{n-2}(x)
\nonumber\\
&&\hspace{-2cm}-\frac{1}{2}n(n-1)(n-2)(n-3)(4n-3)H_{n-4}(x)\nonumber\\
&&\hspace{-2cm}-\frac{1}{3}n(n-1)(n-2)(n-3)(n-4)(n-5)H_{n-6}(x).
\end{eqnarray}

Substituting this result back into (\ref{eq:seriessol1}) and comparing
coefficients, we find that
\begin{eqnarray}
Q_{n}(x) &=&-\frac{1}{1152}H_{n+6}(x)-\frac{1}{128}(4n+7)H_{n+4}(x) \nonumber \\
&&\quad - \frac{1}{32}\left(7n^2+33n+27\right)H_{n+2}(x)\nonumber \\
&&\quad -\frac{1}{8}n(n-1)\left(7n^2-19n+1\right)H_{n-2}(x) \nonumber \\
&&\quad - \frac{1}{8}n(n-1)(n-2)(n-3)(4n-3)H_{n-4}(x)\nonumber \\
&&\quad - \frac{1}{18}n(n-1)(n-2)(n-3)(n-4)(n-5)H_{n-6}(x),
\end{eqnarray}
and
\[E_n=n+\frac{1}{2}+\frac{1}{8}(30n^2+30n+11)\epsilon^2+\mathcal{O}(\epsilon^3).
\]

Having found expressions for the eigenfunctions of the Hamiltonian, we must
verify that they are $\mathcal{PT}$-normalized to this order in perturbation
theory: $\int_{- \infty}^{\infty}dx\left[\phi_n(x)\right]^2=(-1)^n+\mathcal
{O}(\epsilon^3)$. This determines the value of $a_n$ in (\ref{e27c}):
\[\frac{a_n^2}{2^n n!\sqrt{\pi}}\left(2^n n!\sqrt{\pi}+\epsilon^2\int_{-\infty}
^{\infty}dx\,e^{-x^2}[P_n(x)]^2 \right)=1.\]
Using (\ref{eq:ppoly}) as well as the orthogonality and normalization conditions
for the Hermite functions, we obtain $1=a_n^2\left[1-\left(\frac{41}{18}n^3+
\frac{41}{12}n^2+\frac{32}{9}n+\frac{29}{24}\right)\epsilon^2\right]$, and hence
\[a_n=1+\frac{1}{144}(2n+1)(82n^2+82n+87)\epsilon^2+\mathcal{O}(\epsilon^4).\]

We must also verify that the energy eigenstates are simultaneously eigenstates
of $\mathcal{PT}$. Note that $H_n(x)$ and $Q_n(x)$ are even in $x$ for even $n$,
and odd in $x$ for odd $n$; $P_n(x)$ has the opposite parity, but it contains an
additional factor of $ i$. Hence, all three polynomials are $\mathcal{P
T}$-symmetric for even $n$, and $\mathcal{PT}$ anti-symmetric for odd $n$.
\begin{equation}
\label{eq:Hsym}
\mathcal{PT}H_n(x)=
\left\{ \begin{array} {cc}
H_n(x) & {\rm if~n~is~even}, \\
-H_n(x) & {\rm if~n~is~odd}.  \end{array} \right.
\end{equation}
The same holds for the prefactor $i^n$, and thus $\phi_n(x)$ is indeed
$\mathcal{PT}$-symmetric for all $n$:
\[ \mathcal{PT}\phi_n(x) = \phi_n(x).\]

\subsection{Calculation of the $\mathcal{C}$ Operator}

We can now construct the $\mathcal{C}$ operator for the $ix^3$ theory to order
$\mathcal{O}(\epsilon)$:
\begin{eqnarray}
\mathcal{C}(x,y)
&=& \sum_{n=0}^{\infty}\phi_n(x)\phi_n(y) \nonumber \\
&=& \frac{1}{\sqrt{\pi}} e^{-\frac{1}{2}(x^2+y^2)}
\sum_{n=0}^{\infty}\frac{(-1)^n}{2^nn!} \biggl(H_n(x)+\epsilon P_n(x)\biggr)
\biggl(H_n(y)+\epsilon P_n(y)\biggr) \nonumber \\
&=& \frac{1}{\sqrt{\pi}} e^{-\frac{1}{2}(x^2+y^2)}
\sum_{n=0}^{\infty}\frac{1}{2^nn!}H_n(x)H_n(-y) \nonumber \\
&&\hspace{-1cm}+\epsilon \left[\frac{1}{\sqrt{\pi}} e^{-\frac{1}{2}(x^2+y^2)}
\sum_{n=0}^{\infty}\frac{1}{2^nn!}P_n(x)H_n(-y)+(x \leftrightarrow y)
\right] +\mathcal{O}(\epsilon^2).
\label{eq:C1}
\end{eqnarray}
To proceed, we must make use of the completeness relation for the Hermite
functions. To that end, we first need to express $P_n(x)$ solely in terms of
$H_n(x)$ and derivatives thereof. A comparison of equations (\ref{eq:x3H}) and
(\ref{eq:ppoly}) shows that
\begin{equation}
\label{eq:PinH}
i P_n(x) = \frac{1}{12}H_{n+3}(x)+H_{n+1}(x)-\frac{1}{3}x^3 H_n(x)+2xnH_n(x)
-3n^2 H_{n-1}(x).
\end{equation}
We now use $H_{n+1}(x)=\left(2x-\frac{d}{dx}\right)H_n(x)$ to obtain
\[ H_{n+3}(x)=\left[-\frac{d^3}{dx^3}+6x\frac{d^2}{dx^2}+6(1-2x^2)\frac{d}{dx}
+4(2x^3-3x) \right] H_n(x).\]
Also, from $2nH_n(x)=\left[-\frac{d^2}{dx^2}+2x\frac{d}{dx}\right]H_n(x)$,
we have
\[ -3n^2H_{n-1}(x) = -\frac{3}{2}nH'_n(x) = \left[\frac{3}{4}\frac{d^3}{dx^3}
-\frac{3}{2}x\frac{d^2}{dx^2} -\frac{3}{2}\frac{d}{dx}\right]H_n(x).\]

Substituting all this into (\ref{eq:PinH}) gives
\[ i P_n(x)=\left[\frac{2}{3}\frac{d^3}{dx^3}-2x\frac{d^2}{dx^2}+(x^2-2)\frac{d}
{dx} +\left(\frac{1}{3}x^3+x \right)\right]H_n(x),\]
which can be rewritten as
\begin{eqnarray}
i P_n(x)=e^{\frac{1}{2}x^2} \Biggl\lbrace
&&\frac{2}{3}\left[\frac{d^3}{dx^3}+3x\frac{d^2}{dx^2}+3(1+x^2)\frac{d}{dx}+
(x^3+3x) \right] \nonumber \\
&&-2x\left(\frac{d^2}{dx^2}+2x\frac{d}{dx}+(1+x^2)\right)\nonumber \\
&&+(x^2-2)\left[x+\frac{d}{dx}\right]+\left(x+\frac{1}{3}x^3\right)\Biggr\rbrace
\left( e^{-\frac{1}{2}x^2}H_n(x)\right).
\end{eqnarray}
Finally, we obtain an expression of the form
\begin{equation}
\label{eq:simpleP}
 i P_n(x)=e^{\frac{1}{2}x^2}\delta_x \left( e^{-\frac{1}{2}x^2}H_n(x)\right),
\end{equation}
where the differential operator $\delta_x$ is given by
\begin{equation}
\delta_x=\left[\frac{2}{3}\frac{d^3}{dx^3}-x^2\frac{d}{dx}-x\right].
\end{equation}

It is now easy to complete our calculation of the $\mathcal{C}$ operator. We
simply substitute our new expression for $P_n(x)$ into equation (\ref{eq:C1}):
\begin{eqnarray}
\mathcal{C}(x,y)=\left[1-i\epsilon\left(\frac{4}{3}\frac{d^3}{dx^3}-2x\frac{d}
{dx}x\right)\right]\delta(x+y).
\label{eq:x3C}
\end{eqnarray}

\subsection{Verification of $\mathcal{C}$}

We can perform several checks to ascertain the correctness of the $\mathcal{C}$
operator in (\ref{eq:x3C}). For example, we can verify that the eigenstates of
$H$ are also eigenstates of $\mathcal{C}$:
\begin{equation}
\int_{-\infty}^{\infty}dy\,\mathcal{C}(x,y)\phi_n(y)=(-1)^n\phi_n(x).
\label{eq:checkC}
\end{equation}
Now, equation (\ref{eq:checkC}) is clearly satisfied to zeroth order in
$\epsilon$:
\[\int_{-\infty}^\infty dy\,\delta(x+y)e^{-\frac{1}{2}y^2}H_n(y)
= (-1)^n  e^{-\frac{1}{2}x^2}H_n(x) \]
because $H_n(-x)=(-1)^n H_n(x)$.

To first order in $\epsilon$, (\ref{eq:checkC}) reads
\begin{eqnarray}
(-1)^n  e^{- \frac{1}{2}x^2}P_n(x)
&=&\int_{-\infty}^\infty dy\,\delta(x+y)e^{- \frac{1}{2}y^2}P_n(y)\nonumber\\
&&\hspace{-2cm}-i\epsilon\int_{-\infty}^\infty dy\left(\frac{4}{3}
\frac{d^3}{dx^3}-2x\frac{d}{dx} x\right)\delta(x+y)e^{-\frac{1}{2}y^2}H_n(y).
\end{eqnarray}
Noting that $P_n(-x)=-(-1)^n P_n(x)$ we can write
\begin{eqnarray}
2 i(-1)^n  e^{-\frac{1}{2}x^2}P_n(x) = &&\frac{4}{3} x^3\int_{-\infty}
^{\infty}dy\,\delta(x+y) e^{-\frac{1}{2}y^2}H_n(y) \\
&&+2x p{x}\int_{-\infty}^{\infty}dy\,\delta(x+y)y e^{-\frac{1}{2}y^2}
H_n(y),
\end{eqnarray}
or
\[
 i P_n(x)
=  e^{\frac{1}{2}x^2}\left[\frac{2}{3} x^3\left( e^{-\frac{1}{2}x^2}
H_n(x)\right) -x p{x}\left(x e^{-\frac{1}{2}x^2}H_n(x)\right)\right],
\]
which agrees with the previous result in (\ref{eq:simpleP}).

\subsection{The $\mathcal{C}$ Operator as an Exponential}

Extending the calculation of the $\mathcal{C}$ operator to second order in
perturbation theory presents no new conceptual difficulties over and above the
ones encountered at first order in perturbation theory. We simply cite the
result given in \cite{rr11}:
\begin{eqnarray}
\label{eq:expC}
C(x,y) &=& \bigg[1 -\epsilon\left(\frac{4}{3} p^3-2xy p\right) \nonumber \\
&&\hspace{-1cm}+\epsilon^2 \left(\frac{8}{9}p^6-\frac{8}{3}xyp^4+(2x^2y^2-12)
p^2\right)\bigg]\delta(x+y)+\mathcal{O}(\epsilon^3),
\end{eqnarray}
where $p=-i\frac{d}{dx}$. The structure of this formula suggests that it might
be rewritten as an exponential:
\begin{equation}
C(x,y)=\exp\left[-\epsilon\left(\frac{4}{3}p^3-2xpx\right)\right]\delta(x+y).
\end{equation}
Observe that the $\mathcal{C}$ operator reduces to $\mathcal{P}$ in the limit
where the parameter $\epsilon$ tends to zero in (\ref{eq:x3C}). Note that the
expression for the parity operator $\mathcal{P}$ is $\mathcal{P}=e^{i\pi a^{
\dagger}a}$, where $a^{\dagger}$ and $a$ represent the standard quantum 
mechanical harmonic oscillator raising and lowering operators, respectively. The
combination $a^{\dagger}a$ represents the number operator. It is interesting
that in exponential form the $\mathcal{C}$ operator is a series in odd powers of
$\epsilon$. (Consult Ref.~\cite{rr11} for the $\epsilon^3$ contribution to
$\mathcal{C}$.)

\section{Perturbative Calculation of the $\mathcal{C}$ Operator for an
$\boldsymbol{ix^2y}$ Theory}
\label{s5}

Let us now apply the techniques of the previous section to a quantum mechanical
theory having two degrees of freedom. Again, the $\mathcal{PT}$-symmetric
Hamiltonian consists of a standard harmonic oscillator interacting with a
complex potential
\begin{equation}
\label{eq:x2yham}
H =-\half\frac{\partial^2}{\partial x^2}-\half\frac{\partial^2}{\partial y^2}
+\half x^2+\half y^2+i\epsilon x^2 y.
\end{equation}
This complex H\'enon-Heiles theory was studied in Ref.~\cite{rr14}.

The chain of reasoning culminating in an explicit expression for the $\mathcal{C
}$ operator is much the same as that in the previous section, allowing us to
concentrate on the essentials. In order to solve the
Schr\"odinger equation
\begin{equation}
\label{eq:x2yschrodinger}
H\phi_{mn}(x,y)=E_{mn}\phi_{mn}(x,y),
\end{equation}
we again resort to perturbative methods. To first order in $\epsilon$ we have
\begin{equation}
\label{eq:x2yphi}
\phi_{mn}(x,y)\sim e^{-\frac{1}{2}(x^2+y^2)}\left[H_m(x)H_n(y)+\epsilon
P_{mn}(x,y)\right]
\end{equation}
and
\begin{equation}
E_{mn}=m+n+1+\epsilon A_{mn}.
\end{equation}

\subsection{Calculation of the Energy Eigenstates and their Energies}

To first order in $\epsilon$, equation (\ref{eq:x2yschrodinger}) becomes
$$ \left[- \frac{1}{2}\frac{\partial^2}{\partial x^2}
-\frac{1}{2}\frac{\partial^2}{\partial y^2}
+x \frac{\partial}{\partial x}
+y \frac{\partial}{\partial y}-m-n \right]P_{mn}(x,y)
= (A_{mn}-  i x^2 y)H_m(x)H_n(y). $$
Rewriting $P_{mn}(x,y)$ as a power series in Hermite polynomials,
\[P_{mn}(x,y)=\sum_{r,s}p_{rs}H_r(x)H_s(y),\]
then yields
\begin{equation}
\label{eq:x2yseriessol}
\sum_{r,s}p_{rs}\left[(r-m)+(s-n)\right]H_r(x)H_s(y)=(A_{mn}-ix^2y)H_m(x)H_n(y).
\end{equation}
We then have
\begin{eqnarray}
x^2yH_m(x)H_n(y) &=& \frac{1}{8}H_{m+2}(x)H_{n+1}(y)
+\frac{1}{4}nH_{m+2}(x)H_{n-1}(y) \nonumber\\
&&\hspace{-3cm}+\frac{1}{2}\left(m+\frac{1}{2}\right)H_m(x)H_{n+1}(y)
+\left(m+\frac{1}{2}\right)nH_m(x)H_{n-1}(y) \nonumber\\
&&\hspace{-3cm}+\frac{1}{2}m(m-1)H_{m-2}(x)H_{n+1}(y)+m(m-1)nH_{m-2}(x)H_{n-1}
(y).
\end{eqnarray}
The right side of this equation does not contain any terms in $H_m(x)H_n(y)$,
allowing us to deduce that $A_{mn}=0$ for all $m$ and $n$. Thus, the energy
does not change to first order in perturbation theory.

A comparison of the coefficients in (\ref{eq:x2yseriessol}) reveals that
\begin{eqnarray}
iP_{mn}(x,y)&=&H_{m+2}(x)\left[\frac{1}{24}H_{n+1}(y)+\frac{1}{4}nH_{n-1}(y)
\right]\nonumber \\
&&+\left(m+\frac{1}{2}\right)H_m(x)\left[\frac{1}{2}H_{n+1}(y)-nH_{n-1}(y)
\right]\nonumber \\
&&-m(m-1)H_{m-2}(x)\left[\frac{1}{2}H_{n+1}(y)+\frac{1}{3}nH_{n-1}(y)\right].
\end{eqnarray}
We can rewrite this equation in the form
\begin{equation}
\label{eq:Px2y}
iP_{mn}(x,y)=e^{\frac{1}{2}(x^2+y^2)}\delta_{xy}
\left[ e^{-\frac{1}{2}(x^2+y^2)}H_m(x)H_n(y)\right],
\end{equation}
where the differential operator $\delta_{xy}$ is given by
\begin{eqnarray}
\delta_{xy}=\frac{2}{3}\frac{\partial^3}{\partial x^2\partial y}-\frac{1}{3}x^2
\frac{\partial}{\partial y}-\frac{2}{3}xy\frac{\partial}{\partial x}-\frac{1}{3}
y.
\end{eqnarray}

\subsection{Calculation of the $\mathcal{C}$ Operator}

Having established the form of equation (\ref{eq:Px2y}), it is now
straightforward to calculate the $\mathcal{C}$ operator:
\begin{eqnarray}
\mathcal{C}(x,x';y,y') &=&\sum_{m=0}^{\infty}\sum_{n=0}^{\infty}
\phi_{mn}(x,y)\phi_{mn}(x',y')\nonumber\\
&&\hspace{-3cm}=\left[1-i\epsilon\left(\delta_{xy}+\delta_{x'y'}\right)\right]
\delta(x+x')\delta(y+y')\nonumber\\
&&\hspace{-3cm}=\left[1-i\epsilon\left(\frac{4}{3}\frac{\partial^3}
{\partial x^2\partial y}+\frac{2}{3}x x'\frac{\partial}{\partial y}
-\frac{4}{3}xy \frac{\partial}{\partial x}\right)
\right]\delta(x+x')\delta(y+y'),
\end{eqnarray}
where the parity operator is $\delta(x+x')\delta(y+y')$.

\section{Perturbative Calculation of the $\mathcal{C}$ Operator for an
$\boldsymbol{ixyz}$ Theory}
\label{s6}

As a third example we consider a quantum mechanical theory with three degrees of
freedom. Once again, the $\mathcal{PT}$-symmetric Hamiltonian consists of a
standard harmonic oscillator part interacting with a complex potential
\begin{equation}
\label{eq:xyzham}
H = -\half\frac{\partial^2}{\partial x^2}-\half\frac{\partial^2}{\partial y^2}
-\half\frac{\partial^2}{\partial z^2} +\half x^2 +\half y^2 +\half z^2
+ i\epsilon xyz.
\end{equation}

We wish to solve the Schr\"odinger equation
\begin{equation}
\label{eq:xyzschrodinger}
H\phi_{klm}(x,y,z) = E_{klm}\phi_{klm}(x,y,z),
\end{equation}
whose eigenfunctions are given by
\begin{eqnarray}
\phi_{klm}(x,y,z)&\sim& e^{-\frac{1}{2}r^2}\Big[H_k(x)H_l(y)H_m(z)\nonumber\\
&&+\epsilon P_{klm}(x,y,z)+\epsilon^2 Q_{klm}(x,y,z)\Big]
\label{eq:xyzphi}
\end{eqnarray}
and whose energies have the form
\[ E_{klm}=n+\frac{3}{2}+\epsilon A_{klm}+\epsilon^2 B_{klm},\]
where we have set $r^2=x^2+y^2+z^2$ and $n=k+l+m$.

\subsection{Energy Eigenstates}

To first order in $\epsilon$ (\ref{eq:xyzschrodinger}) becomes
\begin{eqnarray}
&&\hspace{-2cm}\left(-\half\frac{\partial^2}{\partial x^2}-\half\frac{
\partial^2}{\partial y^2}-\half\frac{\partial^2}{\partial z^2}+x\frac{\partial}
{\partial x}+y\frac{\partial}{\partial y}+z\frac{\partial}{\partial z}
-n\right)P_{klm}(x,y,z)\nonumber \\
&=&(A_{klm}- i xyz)H_k(x)H_l(y)H_m(z).
\end{eqnarray}
Rewriting $P_{klm}(x,y,z)$ as a sum of Hermite polynomials,
\[P_{klm}(x,y,z)=\sum_{r,s,t}p_{rst}H_r(x)H_s(y)H_t(z),\]
we obtain the equation
\begin{eqnarray}
&&\sum_{r,s,t}p_{rst}(r+s+t-n)H_r(x)H_s(y)H_t(z)\nonumber\\
&&\qquad=(A_{klm}-ixyz)H_k(x)H_l(y)H_m(z).
\label{eq:xyzseriessol}
\end{eqnarray}

Also, we have
\begin{eqnarray}
xyzH_k(x)H_l(y)H_m(z)&=&\frac{1}{8}H_{k+1}(x)H_{l+1}(y)H_{m+1}(z)\nonumber\\
&&\hspace{-4cm}+\frac{1}{4}\Big[k H_{k-1}(x)H_{l+1}(y)H_{m+1}(z)
+lH_{k+1}(x)H_{l-1}(y)H_{m+1}(z) \nonumber\\
&&\hspace{-4cm} +mH_{k+1}(x)H_{l+1}(y)H_{m-1}(z)\Big]
+\frac{1}{2}\Big[ kl H_{k-1}(x)H_{l-1}(y)H_{m+1}(z)\nonumber\\
&&\hspace{-4cm} +km H_{k-1}(x)H_{l+1}(y)H_{m-1}(z)
+lmH_{k+1}(x)H_{l-1}(y)H_{m-1}(z)\Big] \nonumber\\
&&\hspace{-4cm}+klmH_{k-1}(x)H_{l-1}(y)H_{m-1}(z).
\end{eqnarray}
The right side of this equation, being devoid of terms in $H_k(x)H_l(y)H_m(z)$,
confirms that the energy is unaltered to this order in perturbation theory. In
other words, $A_{klm}=0$ for all $k,l,m$. Comparing coefficients, we then find
that
\begin{eqnarray}
iP_{klm}(x,y,z)&=& \frac{1}{24}H_{k+1}(x)H_{l+1}(y)H_{m+1}(z) \nonumber \\
&&\hspace{-2cm}+\frac{1}{4}\Big[kH_{k-1}(x)H_{l+1}(y)H_{m+1}(z)+lH_{k+1}(x)
H_{l-1}(y)H_{m+1}(z)\nonumber \\
&& \hspace{-2cm} +mH_{k+1}(x)H_{l+1}(y)H_{m-1}(z)\Big]
-\frac{1}{2}\Big[ klH_{k-1}(x)H_{l-1}(y)H_{m+1}(z) \nonumber \\
&& \hspace{-2cm}+kmH_{k-1}(x)H_{l+1}(y)H_{m-1}(z)
+lmH_{k+1}(x)H_{l-1}(y)H_{m-1}(z)\Big]\nonumber \\
&&\hspace{-2cm} -\frac{1}{3}klmH_{k-1}(x)H_{l-1}(y)H_{m-1}(z).
\label{eq:xyzppoly}
\end{eqnarray}

Having established the form of $P_{klm}(x,y,z)$, we now must ensure that the
wavefunctions (\ref{eq:xyzphi}) are correctly $\mathcal{PT}$-normalized to order
$\epsilon$ in the sense that
\begin{equation}
\int\!\!\int\!\!\int dx\,dy\,dz\left[\phi_{klm}(x,y,z)\right]^2 = (-1)^{n}.
\end{equation}
Note that $P_{klm}(xyz)$ does not contain a term in $H_k(x)H_l(y)H_m(z)$. Hence,
from the orthogonality and normalization conditions for the Hermite functions,
it follows that the correctly normalized wavefunction must take the form
\[\phi_{klm}(x,y,z) = \frac{ i^{n}}{\sqrt{\pi^{3/2} 2^{n} k! l! m!}}
e^{-\frac{1}{2}r^2}\left[H_k(x)H_l(y)H_m(z)+\epsilon P_{klm}(x,y,z)\right].\]

\subsection{The $\mathcal{C}$ Operator}

Once again, we must express $P_{klm}(x,y,z)$ solely in terms of $H_k(x)H_l(y)
H_m(z)$ and derivatives thereof before we can apply the standard completeness
relation for the Hermite functions. The result is
\begin{eqnarray}
iP_{klm}(x,y,z)
&=& \Bigg[\frac{2}{3}\frac{\partial^3}{\partial x\partial y\partial z}
-\frac{2}{3}\left(x\frac{\partial^2}{\partial y\partial z}
+y\frac{\partial^2}{\partial x\partial z}
+z\frac{\partial^2}{\partial x\partial y}\right)  \nonumber \\
&&\hspace{-3cm}\qquad \quad +\frac{1}{3}\left(xy\frac{\partial}{\partial z}
+xz\frac{\partial}{\partial y} +yz\frac{\partial}{\partial x}\right)
+\frac{1}{3}xyz \Bigg]H_k(x)H_l(y)H_m(z).
\end{eqnarray}

An equivalent form of the polynomial, but one that is more
amenable to the task in hand, may be obtained from
\begin{eqnarray}
\label{eq:group1}
H_{n+1}(x) &=& e^{\frac{1}{2}x^2} \left[x-\frac{d}{dx}\right]
\left( e^{-\frac{1}{2}x^2}H_n(x)\right), \nonumber\\
nH_{n-1}(x) &=& \frac{1}{2} e^{\frac{1}{2}x^2} \left[x+\frac{d}{dx}\right]
\left( e^{-\frac{1}{2}x^2}H_n(x)\right).
\end{eqnarray}
We can thus write $ i P_{klm}(x,y,z)$ in the compact form
\begin{equation}
\label{eq:simplexyzP}
i P_{klm}(x,y,z) =  e^{\frac{1}{2}r^2}\delta_{xyz}
\left[e^{-\frac{1}{2}r^2}H_k(x)H_l(y)H_m(z)\right],
\end{equation}
where the differential operator $\delta_{xyz}$ is given by
\begin{equation*}
\delta_{xyz}=\frac{2}{3}\frac{\partial^3}{\partial x\partial y\partial z}
-\frac{1}{3}\left(xy\frac{\partial}{\partial z}
+xz\frac{\partial}{\partial y}+yz\frac{\partial}{\partial x}\right).
\end{equation*}
Given the symmetric nature of the operator $\delta_{xyz}$ it is now particularly
easy to derive the form of the $\mathcal{C}$ operator:
\begin{eqnarray}
\mathcal{C}(x,x';y,y';z,z') &=&\sum_{k=0}^{\infty}\sum_{l=0}^{\infty}
\sum_{m=0}^{\infty}\phi_{klm}(x,y,z)\phi_{klm}(x',y',z') \nonumber \\
&&\hspace{-6cm}=\left[1-i\epsilon\left(\delta_{xyz}+\delta_{x'y'z'}\right)
\right]\delta(x+x')\delta(y+y')\delta(z+z')\nonumber\\
&&\hspace{-6cm}=\left(1-2i\epsilon\delta_{xyz}\right)\delta(x+x')\delta(y+y')
\delta(z+z')\nonumber\\
&&\hspace{-6cm}=\Bigg\{1-i\epsilon\Bigg[\frac{4}{3}\frac{\partial^3}{\partial x
\partial y\partial z}-\frac{2}{3}\left(xy\frac{\partial}{\partial z}+xz\frac{
\partial}{\partial y} +yz\frac{\partial}{\partial x}\right)\Bigg]\Bigg\}
\delta(x+x')\delta(y+y')\delta(z+z').
\end{eqnarray}

\subsection{Verification of $\mathcal{C}$}

It is important to verify that the eigenstates of the Hamiltonian are also
eigenstates of the $\mathcal{C}$ operator:
\begin{equation}
\label{eq:checkxyzC}
(-1)^{n}\phi_{klm}(x,y,z)=\int\!\!\!\int\!\!\!\int\!\!
dx'\, dy'\, dz'\,\mathcal{C}(x,x';y,y';z,z')\phi_{klm}(x',y',z').
\end{equation}
To demonstrate this to first order in $\epsilon$, we have
\begin{eqnarray}
(-1)^{n} e^{-\frac{1}{2}r^2}P_{klm}(x,y,z) &=&\int\!\!\!\int\!\!\!\int\!\!
dx'\,dy'\,dz'\,\delta_{x,x'}\delta_{y,y'}\delta_{z,z'}
e^{-\frac{1}{2}r'^2}P_{klm}(x',y',z') \nonumber\\
&&\hspace{-5cm}-2i\delta_{xyz}\int\!\!\!\int\!\!\!\int\!\! dx'\,dy'\,dz'\,\delta
_{x,x'}\delta_{y,y'}\delta_{z,z'}e^{-\frac{1}{2}r'^2}H_k(x')H_l(y')H_m(z'),
\end{eqnarray}
where we have used the abbreviated notation $\delta_{x,x'}$ for $\delta(x+x')$.

Observing that $P_{klm}(-x,-y,-z) = -(-1)^{n}P_{klm}(x,y,z)$, we can write
\[ 2(-1)^{n} e^{-\frac{1}{2}r^2}P_{klm}(x,y,z)=-2i\delta_{xyz}
\left[(-1)^{n} e^{-\frac{1}{2}r^2}H_k(x)H_l(y)H_m(z)\right], \]
whence we recover equation (\ref{eq:simplexyzP}):
\[iP_{klm}(x,y,z)=e^{\frac{1}{2}r^2}\delta(x,y,z)\left[e^{-\frac{1}{2}r^2}H_k(x)
H_l(y)H_m(z)\right]. \]

\subsection{Second Order Perturbation Theory -- Difficulties with Degeneracy}

At order $\epsilon^2$ a tough problem surfaces, namely that of degeneracy. To
second order the eigenproblem (\ref{eq:xyzschrodinger}) becomes
\begin{eqnarray}
\sum_{r,s,t}q_{rst}(r+s+t-n)H_r(x)H_s(y)H_t(z)\nonumber\\
&& \hspace{-4cm} =B_{n}H_k(x)H_l(y)H_m(z)- i xyzP_{klm}(x,y,z),
\label{eq:xyzseriessol2}
\end{eqnarray}
where we posed
\begin{equation}
Q_{klm}(x,y,z) = \sum_{r,s,t} q_{rst} H_r(x)H_s(y)H_t(z).
\end{equation}

In order to find the coefficients $q_{rst}$, we need to express $ixyzP_{klm}(x,
y,z)$ in terms of Hermite polynomials. We use the formula
\[ xH_{n+1}(x)\equiv\half H_{n+2}(x)+(n+1)H_n(x) \]
applied to (\ref{eq:xyzppoly}). The result is a highly symmetric formula that is
too long to give here.

When we now examine (\ref{eq:xyzseriessol2}) in the light of the formula for
$ixyzP_{klm}(x,y,z)$, it soon becomes apparent that we run into an unforseen
problem: the left-hand side of (\ref{eq:xyzseriessol2}) is zero whenever $r+s+t=
n$, but at the same time we have terms like \mbox{$k(k-1)H_{k-2}(x)H_{l}(y)H_{m+
2}(z)$} and permutations thereof on the right-hand side which are clearly not
zero. The underlying cause of this mismatch lies in the symmetric nature of the
Hamiltonian and the associated degeneracy of its eigenvalues. In fact, the
unperturbed eigenvalues $E_n^{(0)}=n+3/2$ are $(n+1)(n+2)/2$-fold degenerate,
where $n$ denotes the energy level. While this degeneracy persists to first
order in perturbation theory, it is partially lifted at second order. As a
result of this degeneracy, one needs to take into consideration a mixing of
states corresponding to the same energy.

We briefly illustrate the technique here by examining the energy levels $n=2$,
4, and 6, which are 6, 15, and 28 fold degenerate, respectively. A little
reflection reveals, however, that not all of these states figure in the mixing.
For the $n=4$ level, for example, we find that we need only include 6 of the 15
states in the mixing. In the study of our three examples we shall need to have
recourse to some special cases of the lengthy equation for $ixyzP_{klm}(x,y,z)$.

\subsection{The $n=2$ Energy Level}

We can repair the inconsistency encountered in (\ref{eq:xyzseriessol2}) by
replacing it here with
\begin{eqnarray}
&& \sum_{r,s,t}q_{rst}\left[(r+s+t-2)\right]H_r(x)H_s(y)H_t(z) \nonumber \\
&& \hspace{-2cm}=B_{2}\bigl[\alpha_1 H_2(x)H_0(y)H_0(z)+\alpha_2 H_0(x)H_2(y)H_0
(z)+\alpha_3 H_0(x)H_0(y)H_2(z)\bigr] \nonumber \\
&& \hspace{-2cm}-i xyz\bigl[\alpha_1 P_{200}(x,y,z)+\alpha_2 P_{020}(x,y,z)
+\alpha_3 P_{002}(x,y,z)\bigr].
\label{eq:eigen2}
\end{eqnarray}
We need to choose the mixing coefficients $\alpha_{1}$, $\alpha_{2}$, and
$\alpha_3$ such that the problem terms $H_2(x)H_0(y)H_0(z)$, $H_0(x)H_2(y)H_0
(z)$, and $H_0(x)H_0(y)H_2(z)$ disappear. This amounts to solving the linear
system of equations
\begin{equation}
\left(
\begin{array}{ccc}
B_2-3/8 & -1/4 & -1/4 \\ -1/4 & B_2-3/8 & -1/4 \\ -1/4 & -1/4 & B_2-3/8
\end{array}
\right)
\left( \begin{array}{c} \alpha_1 \\ \alpha_2 \\ \alpha_3 \end{array}
\right) = \left( \begin{array}{c} 0 \\ 0 \\ 0 \end{array}
\right).
\end{equation}

Cramar's rule states that for a nontrivial solution to exist we must require
that the determinant of the given matrix be zero:
\[ (B_2-1/8)^2 (B_2- 7/8)=0. \]
We see that the effect of the second order contribution to the unperturbed
energy level (of value 3.5 and 3-fold degenerate with respect to the states
under consideration) is to split it into two levels, both of which are raised
and one of which is doubly degenerate. To the doubly degenerate energy there
corresponds the following condition on the mixing coefficients:
\[ \alpha_1+\alpha_2+\alpha_3=0,\quad(\rm{e.g.}\;\alpha_1=1,\,\alpha_2=\alpha_3=
-1/2). \]
This reflects an anti-symmetric mixing of states. For the nondegenerate energy,
the condition reads:
\[\alpha_1=\alpha_2=\alpha_3,\quad(\rm{e.g.}\;\alpha_1=\alpha_2=\alpha_3=1),\]
indicating a symmetric mixing of states.

Let us consider the case of the symmetric mixing of states for illustrative
purposes. Equation (\ref{eq:eigen2}) becomes
\begin{eqnarray}
&& \sum_{r,s,t} q_{rst}\left[(r+s+t-2)\right]H_r(x)H_s(y)H_t(z) \nonumber \\
&& = \frac{7}{8}\bigl[H_2(x)H_0(y)H_0(z)+H_0(x)H_2(y)H_0(z)+H_0(x)H_0(y)H_2(z)
\bigr] \nonumber \\
&& \quad -  i xyz\bigl[P_{200}(x,y,z)+P_{020}(x,y,z)+P_{002}(x,y,z)\bigr].
\end{eqnarray}
From (\ref{eq:xyzppoly}) we know that
\begin{eqnarray}
&& i \bigl[P_{200}(x,y,z)+P_{020}(x,y,z)+P_{002}(x,y,z)\bigr]\nonumber \\
=&& \frac{1}{24}\bigl[H_{3}(x)H_{1}(y)H_{1}(z)+ H_{1}(x)H_{3}(y)H_{1}(z)+
H_{1}(x)H_{1}(y)H_{3}(z)\bigr]\nonumber \\
&&+ \frac{3}{2}H_{1}(x)H_{1}(y)H_{1}(z),
\end{eqnarray}
and in addition, we have
\begin{eqnarray}
&& i xyz\left[P_{200}(x,y,z)+P_{020}(x,y,z)+P_{002}(x,y,z)\right] \nonumber\\
&&\hspace{-1cm}= \frac{1}{192}\Bigl[H_{4}(x)H_{2}(y)H_{2}(z)+
H_{4}(x)H_{2}(y)H_{4}(z)+H_{2}(x)H_{2}(y)H_{4}(z)\Bigr]\nonumber \\
&& \hspace{-1cm}+\frac{1}{48}\Bigl[H_{4}(x)H_{0}(y)H_{0}(z)+H_{0}(x)H_{4}(y)
H_{0}(z)+ H_{0}(x)H_{0}(y)H_{4}(z)\Bigr]\nonumber \\
&& \hspace{-1cm}+\frac{1}{2}\Bigl[H_{2}(x)H_{2}(y)H_{0}(z)+ H_{2}(x)H_{0}(y)
H_{2}(z)+H_{0}(x)H_{2}(y)H_{2}(z)\Bigr]\nonumber \\
&& \hspace{-1cm}+\frac{7}{8}\Bigl[H_{2}(x)H_{0}(y)H_{0}(z)+H_{0}(x)H_{2}(y)
H_{0}(z)+H_{0}(x)H_{0}(y)H_{2}(z)\Bigr]\nonumber \\
&& \hspace{-1cm}+\frac{1}{96}\Big[H_{4}(x)H_{2}(y)H_{0}(z)+H_{4}(x)H_{0}(y)
H_{2}(z)\nonumber\\
&& \hspace{-1cm} +H_{2}(x)H_{4}(y)H_{0}(z)+H_{2}(x)H_{0}(y)H_{4}(z)\nonumber \\
&& \hspace{-1cm} +H_{0}(x)H_{4}(y)H_{2}(z)+H_{0}(x)H_{2}(y)H_{4}(z)\Big]
\nonumber \\
&&\hspace{-1cm}+\frac{9}{32}H_{2}(x)H_{2}(y)H_{2}(z)+\frac{3}{2}H_0(x)H_0(y)
H_0(z).
\end{eqnarray}
Having done this analysis, we now observe that the problem terms $H_2(x)H_0(y)
H_0(z)$, $H_0(x)H_2(y)H_0(z)$, and $H_0(x)H_0(y)H_2(z)$ do indeed disappear.

Finally, we can determine the coefficients $q_{rst}$ in
(\ref{eq:eigen2}). We find that
\begin{eqnarray}
&&Q_{200}(x,y,z)+Q_{020}(x,y,z)+Q_{002}(x,y,z) \nonumber\\
&&\hspace{-2cm}=-\frac{1}{1152}\bigl[H_4(x)H_2(y)H_2(z)+H_2(x)H_4(y)H_2(z)
+ H_{2}(x)H_{2}(y)H_{4}(z)\bigr] \nonumber\\
&&\hspace{-2cm} -\frac{1}{96}\bigl[H_4(x)H_0(y)H_0(z)+H_0(x)H_4(y)H_0(z)+
H_0(x)H_0(y)H_4(z)\bigr] \nonumber\\
&&\hspace{-2cm}-\frac{1}{4}\bigl[H_2(x)H_2(y)H_0(z)+H_2(x)H_0(y)H_2(z)+
H_0(x)H_2(y)H_2(z)\bigr] \nonumber\\
&&\hspace{-2cm}-\frac{1}{384}\Big[H_4(x)H_2(y)H_0(z)+H_4(x)H_0(y)H_2(z)
\nonumber\\
&&\hspace{-2cm}+H_{2}(x)H_{4}(y)H_{0}(z)+H_{2}(x)H_{0}(y)H_{4}(z) \nonumber\\
&&\hspace{-2cm}+H_0(x)H_4(y)H_2(z)+H_0(x)H_2(y)H_4(z)\Big]\nonumber\\
&&\hspace{-2cm}-\frac{9}{128}H_2(x)H_2(y)H_2(z)+\frac{3}{4}H_0(x)H_0(y)H_0(z).
\end{eqnarray}

\subsection{The $n=4$ Energy Level}

At this energy level we assume a structure of the form
\begin{eqnarray}
&& \sum_{r,s,t} q_{rst}\left[(r+s+t-4)\right]H_r(x)H_s(y)H_t(z)\nonumber \\
&& \hspace{-1cm}
= B_{4}\Big[ \alpha_{1}H_4(x)H_0(y)H_0(z)+ \alpha_{2}H_0(x)H_4(y)H_0(z)+
\alpha_{3}H_0(x)H_0(y)H_4(z)\nonumber \\
&& \hspace{-1cm}
+ \beta_{1} H_0(x)H_2(y)H_2(z)+
\beta_{2} H_2(x)H_0(y)H_2(z)+
\beta_{3} H_2(x)H_2(y)H_0(z)\Big]\nonumber \\
&& \hspace{-1cm}
-  i xyz\Big[ \alpha_{1}P_{400}(x,y,z)+
\alpha_{2}P_{040}(x,y,z)+\alpha_{3}P_{004}(x,y,z)\nonumber \\
+&& \hspace{-1cm}\beta_{1}P_{022}(x,y,z)+\beta_{2}P_{202}(x,y,z)
+\beta_{3}P_{220}(x,y,z)\Big].
\end{eqnarray}
Now, we need to choose the mixing coefficients $\alpha_1$, $\alpha_2$,
$\alpha_3$, $\beta_1$, $\beta_2$, and $\beta_3$ such that the six problem terms
disappear. Equivalently, we need to solve the system of linear equations
$\mathbf{M}.\mathbf{v}=\mathbf{0}$, where the coefficient matrix $\mathbf{M}$ of
the system is given by
\begin{equation}
\mathbf{M}=\left[\begin{array}{cccccc}
B_{4}- \frac{17}{24}&0&0&0&- \frac{1}{4}&- \frac{1}{4} \\
0&B_4- \frac{17}{24}&0&- \frac{1}{4}&0&- \frac{1}{4} \\
0&0&B_{4}- \frac{17}{24}&- \frac{1}{4}&- \frac{1}{4}&0 \\
0&- \frac{3}{2}&- \frac{3}{2}&B_{4}- \frac{11}{8}&- \frac{1}{4}&- \frac{1}{4} \\
- \frac{3}{2}&0&- \frac{3}{2}&- \frac{1}{4}&B_{4}- \frac{11}{8}& - \frac{1}{4}\\
- \frac{3}{2}&- \frac{3}{2}&0&- \frac{1}{4}&- \frac{1}{4}&B_{4}- \frac{11}{8} \\
\end{array}
\right]
\end{equation}
and $\mathbf{v}$ denotes the column vector $(\alpha_1,\alpha_2,\alpha_3,\beta_1,
\beta_2,\beta_3)$.

For a nontrivial solution we require that the determinant of the
given matrix be zero, so
\[(192B_4^2-496B_4-33)(192B_4^2-352B_4+81)^2=0.\]
The unperturbed energy level (of value 5.5 and 6-fold degenerate) is split into
four distinct levels (3 raised, 1 lowered), of which two are doubly degenerate.

Associated with the doubly degenerate values of the energy are the
following conditions on the mixing coefficients:
\[ \alpha_1+\alpha_2+\alpha_3=0~{\rm and}~\beta_1+\beta_2+\beta_3=0,~{\rm
with}~\beta_1=\frac{1}{6}(-5\mp\sqrt{241})\alpha_1. \]
Hence, this case corresponds to an anti-symmetric mixing of states.

For the nondegenerate energies, one has
\[\alpha_1=\alpha_2=\alpha_3~{\rm and}~\beta_1=\beta_2=\beta_3,~{\rm with}~\beta
_1=\frac{1}{6}(7\pm\sqrt{265})\alpha_1,\]
which yields a symmetric mixing of states. We observe that the unperturbed
energy level (of value 7.5 and 10-fold degenerate) is split it into seven levels
(6 raised, 1 lowered), of which three are doubly degenerate.

To $B_{6}= \frac{5}{8}$ there correspond the conditions
\begin{eqnarray}
\alpha_1=\alpha_2=\alpha_3\equiv 0,~\beta_1=\beta_3=\beta_5,~\beta_2=\beta_4=
\beta_6,\nonumber\\
\beta_1+\beta_2+\beta_3+\beta_4+\beta_5+\beta_6\equiv 0,~\gamma\equiv0.\nonumber
\end{eqnarray}
We find that an odd permutations of the indices of the states introduces a
relative minus sign.

The symmetry of the states associated with the nondegenerate energies ($B_{6}=
5.473, 2.343, 0.391$) is characterized by
\[\alpha_1=\alpha_2=\alpha_3,~\beta_1=\beta_2=\beta_3=\beta_4=\beta_5=\beta_6,\]
with the relations
\begin{eqnarray}
\beta_1 = 8.863 \alpha_1 \; {\rm and}\; \gamma=3.701 \beta_1=32.804 \alpha_1,
\nonumber\\
\beta_1=2.603\alpha_1\;{\rm and}\;\gamma=-12.888 \beta_1=-33.553\alpha_1,
\nonumber\\
\beta_1=-1.300\alpha_1\; {\rm and}\; \gamma=-3.396\beta_1=4.415\alpha_1,
\nonumber
\end{eqnarray}
respectively. Finally, for the case $n=6$ the states corresponding to the doubly
degenerate energies ($B_{6}=4.003, 1.981, -0.193$) mix according to
\[\alpha_1+\alpha_2+\alpha_3\equiv 0,~\beta_1+\beta_2+\beta_3+\beta_4+\beta_5+
\beta_6 \equiv 0,~\gamma\equiv0.\]

On the basis of the three cases considered, one can see that there is a direct
correlation between the degeneracy of an energy level and the mixing symmetry of
its associated state. An anti-symmetric mixing of states corresponds to a doubly
degenerate eigenvalue, while a symmetric mixing of states or a state of mixed
symmetry corresponds to a nondegenerate energy. Moreover, the number of (not
necessarily distinct) energies equals the number of states we are mixing (in our
cases: 3, 6, and 10).

In conclusion, it is necessary to take great care to deal with the difficulties
presented by degeneracies. Through the examination of three examples we have
found that the states corresponding to a degenerate energy mix according to
certain symmetry criteria. These clearly have to be respected when one is
attempting to calculate the $\mathcal{C}$ operator. Obviously, the problem of
degenerate states makes it very difficult to calculate the $\mathcal{C}$
operator in systems having more that one degree of freedom. The problems
associated with degeneracy can, in fact, be overcome and the techniques for
doing so are described in a paper under preparation by Bender, Brody, and Jones
\cite{rr15}; in this paper it is shown that it is even possible to find
$\mathcal{C}$ for systems having an infinite number of degrees of freedom
(quantum field theory).

\section{Applications and Possible Observable Consequences}
\label{s7}

We do not know if non-Hermitian, $\mathcal{PT}$-symmetric Hamiltonians can be
used to describe experimentally observable phenomena. However, non-Hermitian
Hamiltonians have {\it already} been used to describe interacting systems. For
example, Wu showed that the ground state of a Bose system of hard spheres is
described by a non-Hermitian Hamiltonian \cite{rr16}. Wu found that the
ground-state energy of this system is real and conjectured that all energy
levels were real. Hollowood showed that even though the Hamiltonian of a complex
Toda lattice is non-Hermitian, its energy levels are real \cite{rr17}.
Non-Hermitian Hamiltonians of the form $H=p^2+ix^3$ also arise in Reggeon field
theory models that exhibit real positive spectra \cite{rr18}. In each of these
cases the fact that a non-Hermitian Hamiltonian had a real spectrum appeared
mysterious at the time, but now the explanation is simple: In each case the
non-Hermitian Hamiltonian is $\mathcal{PT}$-symmetric and in each case the
Hamiltonian was constructed so that the position operator $x$ or the field
operator $\phi$ is always multiplied by $i$.

An experimental signal of a complex Hamiltonian might be found in the context of
condensed matter physics. Consider the complex crystal lattice whose potential
is $V(x)=i\sin\,x$. While the Hamiltonian $H=p^2+i\sin\,x$ is not Hermitian, it
is $\mathcal{PT}$-symmetric and all of its energy bands are {\it real}.
However, at the edge of the bands the wave function of a particle in such a
lattice is always bosonic ($2\pi$-periodic) and, unlike the case of ordinary
crystal lattices, the wave function is never fermionic ($4\pi$-periodic)
\cite{rr19}. Direct observation of such a band structure would give unambiguous
evidence of a $\mathcal{PT}$-symmetric Hamiltonian.

There are many opportunities for the use of non-Hermitian Hamiltonians in the
study of quantum field theory. Many field theory models whose Hamiltonians are
non-Hermitian and $\mathcal{PT}$-symmetric have been studied: $\mathcal{P
T}$-symmetric electrodynamics is a particularly interesting theory because it is
asymptotically free (unlike ordinary electrodynamics) and because the direction
of the Casimir force is the negative of that in ordinary electrodynamics
\cite{rr20}. This theory is remarkable because it can determine its own coupling
constant. Supersymmetric $\mathcal{PT}$-symmetric quantum field theories have
also been studied \cite{rr21}. A scalar quantum field theory with a cubic
self-interaction described by the Lagrangian $\mathcal{L}=\half\pi^2+\half
(\nabla\varphi)^2+\half\mu^2\varphi^2+g\varphi^3$ is physically unacceptable
because the energy spectrum is not bounded below. However, the cubic scalar
quantum field theory that corresponds to $H$ in (\ref{e1}) with $\epsilon=1$ is
given by the Lagrangian density
$$\mathcal{L}=\half\pi^2+\half(\nabla\varphi)^2+\half\mu^2\varphi^2+ig\varphi^3.
$$
This is a new, physically acceptable quantum field theory.

We have found that $\mathcal{PT}$-symmetric quantum field theories exhibit
surprising and new phenomena. For example, consider the theory that corresponds
to $H$ in (\ref{e1}) with $\epsilon=2$, which is described by the Lagrangian
density
\begin{equation}
\mathcal{L}=\half\pi^2+\half(\nabla\varphi)^2+\half\mu^2\varphi^2-{1\over4}g
\varphi^4.
\label{e28}
\end{equation}
For example, for $g$ sufficiently small this $-g\varphi^4$ theory possesses
bound states (the conventional $g\varphi^4$ theory does not because the
potential is repulsive). The bound states occur for all dimensions $0\leq D<3$
\cite{rr22}, but for purposes of illustration we describe the bound states in
the context of one-dimensional quantum field theory (quantum mechanics). For the
conventional anharmonic oscillator, which is described by the Hamiltonian
\begin{equation}
H={1\over2}p^2+{1\over2}m^2x^2+{1\over4}gx^4\qquad(g>0),
\label{e29}
\end{equation}
the small-$g$ Rayleigh-Schr\"odinger perturbation series for the $k$th energy
level $E_k$ is
\begin{eqnarray}
E_k\sim m\left[k+{1\over2}+{3\over4}(2k^2+2k+1)\nu+{\rm O}(\nu^2)\right]
\qquad(\nu\to0^+),
\label{e30}
\end{eqnarray}
where $\nu=g/(4m^3)$. The {\it renormalized mass} $M$ is defined as the first
excitation above the ground state: $M\equiv E_1-E_0\sim m[1+3\nu+{\rm O}(
\nu^2)]$ as $\nu\to0^+$.

To determine if the two-particle state is bound, we examine the second
excitation above the ground state using (\ref{e30}). We define
\begin{eqnarray}
B_2\equiv E_2-E_0\sim m\left[2+9\nu+{\rm O}(\nu^2)\right]\qquad(\nu\to0^+).
\label{e31}
\end{eqnarray}
If $B_2<2M$, then a two-particle bound state exists and the (negative) binding
energy is $B_2-2M$. If $B_2>2M$, then the second excitation above the vacuum is
interpreted as an unbound two-particle state. We see from (\ref{e31}) that in
the small-coupling region, where perturbation theory is valid, the conventional
anharmonic oscillator does not possess a bound state. Indeed, using WKB,
variational methods, or numerical calculations, one can show that there is no
two-particle bound state for any value of $g>0$. Because there is no bound state
the $gx^4$ interaction may be considered to represent a repulsive force. Note
that in general, a repulsive force in a quantum field theory is represented by
an energy dependence in which the energy of a two-particle state decreases with
separation. The conventional anharmonic oscillator Hamiltonian corresponds to a
field theory in one space-time dimension, where there cannot be any spatial
dependence. In this case the repulsive nature of the force is understood to mean
that the energy $B_2$ needed to create two particles at a given time is more
than twice the energy $M$ needed to create one particle.

The perturbation series for the non-Hermitian, $\mathcal{PT}$-symmetric
Hamiltonian
\begin{equation}
H={1\over2}p^2+{1\over2}m^2x^2-{1\over4}gx^4\qquad(g>0),
\label{e32}
\end{equation}
is obtained from the perturbation series for the conventional anharmonic
oscillator by replacing $\nu\to-\nu$. Thus, while the conventional anharmonic
oscillator does not possess a two-particle bound state, the $\mathcal{P
T}$-symmetric oscillator does indeed possess such a state. We measure the
binding energy of this state in units of the renormalized mass $M$ and we define
the {\it dimensionless} binding energy $\Delta_2$ by
\begin{eqnarray}
\Delta_2\equiv(B_2-2M)/M\sim-3\nu+{\rm O}(\nu^2)\qquad(\nu\to0^+).
\label{e33}
\end{eqnarray}
This bound state disappears when $\nu$ increases beyond $\nu=0.0465$. As $\nu$
continues to increase, $\Delta_2$ reaches a maximum value of $0.427$ at $\nu=
0.13$ and then approaches the limiting value $0.28$ as $\nu\to\infty$.

In the $\mathcal{PT}$-symmetric anharmonic oscillator, there are not only
two-particle bound states for small coupling constant but also $k$-particle
bound states for all $k\geq2$. The dimensionless binding energies are
\begin{eqnarray}
\Delta_k\equiv(B_k-kM)/M\sim-3k(k-1)\nu/2+{\rm O}(\nu^2)\qquad(\nu\to0+).
\label{e34}
\end{eqnarray}
The coefficient of $\nu$ is negative. Since the dimensionless binding energy
becomes negative as $\nu$ increases from $0$, there is a $k$-particle bound
state. The higher multiparticle bound states cease to be bound for smaller
values of $\nu$; starting with the three-particle bound state, the binding
energy of these states becomes positive as $\nu$ increases past $0.039$,
$0.034$, $0.030$, and $0.027$.

Thus, for any value of $\nu$ there are always a finite number of bound states
and an infinite number of unbound states. The number of bound states decreases
with increasing $\nu$ until there are no bound states at all. There is a range
of $\nu$ for which there are only two- and three-particle bound states, just
like the physical world in which one observes only states of two and three bound
quarks. In this range of $\nu$ if one has an initial state containing a number
of particles (renormalized masses), these particles will clump together into
bound states, releasing energy in the process. Depending on the value of $\nu$,
the final state will consist either of two- or of three-particle bound states,
whichever is energetically favored. There is a special value of $\nu$ for which
two- and three-particle bound states can exist in thermodynamic equilibrium.

How does a $g\varphi^3$ theory compare with a $g\varphi^4$ theory? A $g\varphi^3
$ theory has an attractive force. Bound states arising as a consequence of this
force can be found by using the Bethe-Salpeter equation. However, the $g 
\varphi^3$ field theory is unacceptable because the spectrum is not bounded
below. If we replace $g$ by $ig$, the spectrum becomes real and positive, but
now the force becomes repulsive and there are no bound states. The same is true
for a two-scalar theory with interaction of the form $ig\varphi^2\chi$, which
is an acceptable model of scalar electrodynamics that has no analog of
positronium.

Another feature of $\mathcal{PT}$-symmetric quantum field theory that
distinguishes it from conventional quantum field theory is the commutation
relation between the $\mathcal{P}$ and $\mathcal{C}$ operators. If we write
$\mathcal{C}=\mathcal{C}_{\rm R}+i\mathcal{C}_{\rm I}$, where $\mathcal{C}_{\rm
R}$ and $\mathcal{C}_{\rm I}$ are real, then $\mathcal{C}_{\rm R}\mathcal{P}=
\mathcal{P}\mathcal{C}_{\rm R}$ and $\mathcal{C}_{\rm I}\mathcal{P}=-\mathcal{P}
\mathcal{C}_{\rm I}$. These commutation and anticommutation relations suggest
the possibility of interpreting $\mathcal{PT}$-symmetric quantum field theory as
describing both bosonic and fermionic degrees of freedom, an idea analogous to
the supersymmetric quantum theories. The distinction here, however, is that the
supersymmetry can be broken; that is, bosonic and fermionic counterparts can
have different masses without breaking the $\mathcal{PT}$ symmetry. Therefore,
another possible observable experimental consequence might be the breaking of
the supersymmetry.

\section{Concluding Remarks}
\label{s8}
We have described an alternative to the axiom of standard quantum mechanics that
the Hamiltonian must be Hermitian. We have shown that Hermiticity may be
replaced by the more physical condition of $\mathcal{PT}$ (space-time
reflection) symmetry. Space-time reflection symmetry is distinct from the
conventional Dirac condition of Hermiticity, so it is possible to consider new
quantum theories, such as quantum field theories whose self-interaction
potentials are $ig \varphi^3$ or $-g\varphi^4$. Such theories have previously
been thought to be mathematically and physically unacceptable because the
spectrum might not be real and because the time evolution might not be unitary.

These new kinds of theories are extensions of ordinary quantum mechanics into
the complex plane; that is, continuations of real symmetric Hamiltonians to
complex Hamiltonians. The idea of analytically continuing a Hamiltonian was
first discussed by Dyson, who argued heuristically that perturbation theory for
quantum electrodynamics diverges \cite{rr23}. Dyson's argument involves rotating
the electric charge $e$ into the complex plane $e\to ie$. Applied to the
anharmonic oscillator (\ref{e29}), Dyson's argument goes as follows: If the
coupling constant $g$ is continued in the complex-$g$ plane to $-g$, then the
potential is no longer bounded below, so the resulting theory has no ground
state. Thus, the ground-state energy $E_0(g)$ has an abrupt transition at $g=0$.
As a series in powers of $g$, $E_0(g)$ must have a zero radius of convergence
because $E_0(g)$ is singular at $g=0$. Hence, the perturbation series must
diverge for all $g\neq0$. The perturbation series does indeed diverge, but this
heuristic argument is flawed because the spectrum of the Hamiltonian (\ref{e32})
that is obtained remains ambiguous until the boundary conditions that the wave
functions must satisfy are specified. The spectrum depends crucially on how this
Hamiltonian with a negative coupling constant is obtained.

There are two ways to obtain $H$ in (\ref{e32}). First, one can substitute $g=|g
|e^{i\theta}$ into (\ref{e29}) and rotate from $\theta=0$ to $\theta=\pi$. Under
this rotation, the ground-state energy $E_0(g)$ becomes complex. Evidently, $E_0
(g)$ is real and positive when $g>0$ and complex when $g<0$. Note that rotating
from $\theta=0$ to $\theta=-\pi$, we obtain the same Hamiltonian as in
(\ref{e32}) but the spectrum is the complex conjugate of the spectrum obtained
when we rotate from $\theta=0$ to $\theta=\pi$. Second, one can obtain
(\ref{e32}) as a limit of the Hamiltonian
\begin{equation}
H={1\over2}p^2+{1\over2}m^2x^2+{1\over4}gx^2(ix)^\epsilon\qquad(g>0)
\label{e35}
\end{equation}
as $\epsilon:0\to2$. The spectrum of this Hamiltonian is real, positive, and
discrete. The spectrum of the limiting Hamiltonian (\ref{e32}) obtained in this
manner is similar in structure to that of the Hamiltonian in (\ref{e29}).

How can the Hamiltonian (\ref{e32}) possess two such astonishingly different
spectra? The answer lies in the boundary conditions satisfied by the wave
functions $\phi_n(x)$. In the first case, in which $\theta={\rm arg}\,g$ is
rotated in the complex-$g$ plane from $0$ to $\pi$, $\psi_n(x)$ vanishes in the
complex-$x$ plane as $|x|\to\infty$ inside the wedges $-\pi/3<{\rm arg}\,x<0$
and $-4\pi/3<{\rm arg}\,x<-\pi$. In the second case, in which the exponent
$\epsilon$ ranges from $0$ to $2$, $\phi_n(x)$ vanishes in the complex-$x$ plane
as $|x|\to\infty$ inside the wedges $-\pi/3<{\rm arg}\,x<0$ and $-\pi<{\rm arg}
\,x<-2\pi/3$. In this second case the boundary conditions hold in wedges that
are symmetric with respect to the imaginary axis; these boundary conditions
enforce the $\mathcal{PT}$ symmetry of $H$ and are responsible for the reality
of the energy spectrum.

Apart from the spectra, there is yet another striking difference between the two
theories corresponding to $H$ in (\ref{e32}). The one-point Green's function
$G_1(g)$ is defined as the expectation value of the operator $x$ in the
ground-state wave function $\phi_0(x)$,
\begin{equation}
G_1(g)=\langle0|x|0\rangle/\langle0|0\rangle\equiv\int_C dx\,x\psi_0^2(x)\Bigm/
\int_C dx\,\psi_0^2(x),
\label{e36}
\end{equation}
where $C$ is a contour that lies in the asymptotic wedges described above. The
value of $G_1(g)$ for $H$ in (\ref{e32}) depends on the limiting process by
which we obtain $H$. If we substitute $g=g_0e^{i\theta}$ into the Hamiltonian
(\ref{e29}) and rotate from $\theta=0$ to $\theta=\pi$, we find by an elementary
symmetry argument that $G_1(g)=0$ for all $g$ on the semicircle in the
complex-$g$ plane. Thus, this rotation in the complex-$g$ plane preserves parity
symmetry ($x\to-x$). However, if we define $H$ in (\ref{e32}) by using the
Hamiltonian in (\ref{e35}) and by allowing $\epsilon$ to range from $0$ to $2$,
we find that $G_1(g)\neq0$. Indeed, $G_1(g)\neq0$ for {\it all} values of
$\epsilon>0$. Thus, in this theory $\mathcal{PT}$ symmetry (reflection about the
imaginary axis, $x\to-x^*$) is preserved, but parity symmetry is permanently
broken.

Finally, we point out that the ``wrong-sign'' field theory described by the
Lagrangian density (\ref{e28}) is remarkable because, in addition to the energy
spectrum being real and positive, the one-point Green's function (the vacuum
expectation value of the field $\varphi$) is {\it nonzero} \cite{rr24}.
Furthermore, the field theory is renormalizable, and in four dimensions is
asymptotically free (and thus nontrivial) \cite{rr25}. Based on these features,
we believe that a $-g\varphi^4$ quantum field theory the theory may provide a
useful setting to describe the dynamics of the Higgs sector in the standard
model.

\section*{Acknowledgements}
CMB is grateful to the Theoretical Physics Group at Imperial College for their
hospitality and he thanks the U.K. Engineering and Physical Sciences Research
Council, the John Simon Guggenheim Foundation, and the U.S.~Department of Energy
for financial support.


\bigskip
\begin{thebibliography}{999}

\bibitem{rr1} Bender C M, Boettcher S, and Meisinger P N 1999 \JMP {\bf 40} 2201

\bibitem{rr2} Bender C M, Meisinger P N, and Wang Q 2003 \JPA {\bf 36} 1029

\bibitem{rr3} Bender C M and Boettcher S 1998 \PRL {\bf 80} 5243. This reference
contains an extended discussion of the wedges in the complex-$x$ plane and the
contour along which the differential equation (\ref{e3}) is solved. See
especially Fig.~2 in this reference.

\bibitem{rr4} Dorey P, Dunning C, and Tateo R 2001 \JPA {\bf 34} L391; Dorey P,
Dunning C, and Tateo R 2001 \JPA {\bf 34} 5679

\bibitem{rr5} L\'evai G and Znojil M 2000 \JPA {\bf 33} 7165; Bagchi B and
Quesne C 2002 \PL A {\bf 300} 18; Trinh D T 2002 PhD Thesis, University of
Nice-Sophia Antipolis, and references therein

\bibitem{rr6} Mostafazadeh A 2002 \JMP {\bf 43} 205; {\em ibid} 2814; {\em ibid}
3944; Ahmed Z 2002 \PL A {\bf 294} 287; Japaridze G S (2002) \JPA {\bf 35} 1709;
Znojil M Preprint (math-ph/0104012); Ramirez A and Mielnik B 2003 {\it Rev. Mex.
Phys.} {\bf 49} 130 

\bibitem{rr7} Bender C M, Brody D C, and Jones H F 2002 \PRL
{\bf 89} 270401 and 2003 {\it Am.~J.~Phys.} {\bf 71} 1095

\bibitem{rr8} Bender C M, Boettcher S, and Savage V M 2000 \JMP {\bf 41} 6381; 
Bender C M, Boettcher S, Meisinger P N, and Wang Q 2002 \PL A {\bf 302} 286

\bibitem{rr9} Mezincescu G A 2000 \JPA {\bf 33} 4911; Bender C M and Wang Q 2001
\JPA {\bf 34} 3325

\bibitem{rr10} Dirac P A M 1942 \PRS {\it London} A {\bf 180} 1

\bibitem{rr11} Bender C M, Meisinger P N, and Wang Q 2003 \JPA {\bf 36} 1973

\bibitem{rr12} Bender C M, Meisinger P N, and Wang Q manuscript in preparation

\bibitem{rr13} Bender C M, Berry M V, and Mandilara A 2002 \JPA {\bf 35} L467

\bibitem{rr14} Bender C M, Dunne G V, Meisinger P N, and \d{S}im\d{s}ek, M 2001
\PL A {\bf 281} 311

\bibitem{rr15} Bender, C M, Brody, D C, and Jones, H F 2004 preprint

\bibitem{rr16} Wu T T 1959 \PR {\bf 115} 1390

\bibitem{rr17} Hollowood T 1992 \NP B {\bf 384} 523

\bibitem{rr18} Brower R, Furman M, and Moshe M 1978 \PL B {\bf 76} 213

\bibitem{rr19} Bender C M, Dunne G V, and Meisinger P N 1999 \PL A {\bf 252} 272

\bibitem{rr20} Bender C M and Milton K A 1999 \JPA {\bf 32} L87

\bibitem{rr21} Bender C M and Milton K A 1998 \PR D {\bf 57} 3595

\bibitem{rr22} Bender C M, Boettcher S, Jones H F, Meisinger P N, and 
\d{S}im\d{s}ek M 2001 \PL A {\bf 291} 197

\bibitem{rr23} Dyson F J 1952 \PR {\bf 85} 631

\bibitem{rr24} Bender C M, Meisinger P N, and Yang H 2001 \PR D {\bf 63} 45001

\bibitem{rr25} Bender C M, Milton K A, and Savage V M 2000 \PR D {\bf 62} 85001

\end{thebibliography}
\end{document}